\documentclass[modern]{aastex631}

\usepackage{amsmath, amssymb, bm}
\usepackage{graphicx}

\DeclareMathOperator*{\argmax}{arg\,max}

\begin{document}

\title{Is there a retrograde accretion disk around 4U 1626$-$67? Tracking torque reversals with a state-space model}

\author{Joseph O'Leary}
\affiliation{School of Physics, University of Melbourne, Parkville, VIC 3010, Australia.}
\affiliation{Australian Research Council Centre of Excellence for Gravitational Wave Discovery (OzGrav), Parkville, VIC 3010, Australia.}

\author{Andrew Melatos}
\affiliation{School of Physics, University of Melbourne, Parkville, VIC 3010, Australia.}
\affiliation{Australian Research Council Centre of Excellence for Gravitational Wave Discovery (OzGrav), Parkville, VIC 3010, Australia.}

\author{Tom Kimpson}
\affiliation{School of Physics, University of Melbourne, Parkville, VIC 3010, Australia.}
\affiliation{Australian Research Council Centre of Excellence for Gravitational Wave Discovery (OzGrav), Parkville, VIC 3010, Australia.}

\author{Dimitris M. Christodoulou}
\affiliation{University of Massachusetts Lowell, Kennedy College of Sciences, Lowell, MA 01854, USA.}
\affiliation{Lowell Centre for Space Science and Technology, Lowell, MA 01854, USA.}

\author{Nicholas J. O'Neill}
\affiliation{School of Physics, University of Melbourne, Parkville, VIC 3010, Australia.}
\affiliation{Australian Research Council Centre of Excellence for Gravitational Wave Discovery (OzGrav), Parkville, VIC 3010, Australia.}

\author{Patrick M. Meyers}
\affiliation{Theoretical Astrophysics Group, California Institute of Technology, Pasadena, CA 91125, USA.}

\author{Sayantan Bhattacharya}
\affiliation{University of Massachusetts Lowell, Kennedy College of Sciences, Lowell, MA 01854, USA.}
\affiliation{Lowell Centre for Space Science and Technology, Lowell, MA 01854, USA.}

\author{Silas G.T. Laycock}
\affiliation{University of Massachusetts Lowell, Kennedy College of Sciences, Lowell, MA 01854, USA.}
\affiliation{Lowell Centre for Space Science and Technology, Lowell, MA 01854, USA.}

\begin{abstract}
X-ray timing studies of the persistent, Galactic, accretion-powered pulsar 4U 1626$-$67 reveal torque reversals, during which the pulse frequency $\nu(t)$ alternates between multiyear episodes of secular acceleration and deceleration, separated by transitions lasting $\lesssim 150 \, \rm{days}$. Here an unscented Kalman filter is applied to track the $\nu(t)$ fluctuations observed in 22.7 years (3340 samples) of publicly available Compton Gamma-Ray Observatory and Fermi Gamma-Ray Space Telescope data to test the canonical picture of magnetocentrifugal accretion for consistency with prograde-prograde and retrograde-prograde accretion disk configurations on either side of the 2008 torque reversal. It is found that the retrograde-prograde model is preferred, with a log Bayes factor equal to  0.44 and maximum a posteriori log likelihood ratio equal to 2.5. The mass accretion rate $Q(t)$ and magnetocentrifugal fastness $\omega(t)$ transition smoothly between episodes of deceleration and acceleration; $Q(t)$ shifts by $\leq 0.34 \, {\rm dex}$ across the reversal, and one measures $\omega(t) \approx 0.25$ and $\omega(t) \approx 0.30$ during deceleration and acceleration, respectively. The angular acceleration $\dot{\Omega}(t)$ satisfies  $-9 \lesssim \dot{\Omega}(t)/(10^{-12} \, \rm{rad \, s^{-2}}) \lesssim -5$ and  $2 \lesssim \dot{\Omega}(t)/(10^{-12} \, \rm{rad \, s^{-2}}) \lesssim 9$ before and after the 2008 reversal, respectively, compared to $\dot{\Omega} \approx -3.0 \times 10^{-12} \, \rm{rad \, s^{-2}}$ before reversal and $\dot{\Omega} \approx 2.5 \times 10^{-12} \, \rm{rad \, s^{-2}}$ after reversal, as inferred from previous long-term X-ray timing and spectral analysis of 4U 1626$-$67.

\end{abstract}

\section{Introduction} \label{Sec:Intro}

Torque reversals are a feature of the rotational histories of many X-ray pulsars; see \cite{Bildsten_1997} and \cite{Malacaria_2020} for comprehensive reviews. They are observed in tens of objects in the Milky Way and Magellanic Clouds during the course of long-term X-ray timing campaigns with instruments aboard the Compton Gamma-Ray Observatory [CGRO; \citep{Bildsten_1997,Finger_1997}], Fermi Gamma-Ray Space Telescope [FGRST; \citep{Meegan_2009,Malacaria_2020}], and other satellites. Torque reversals occur, when the star alternates between secular episodes of acceleration and deceleration lasting $\gtrsim 10 \, \rm{yr}$, separated by rapid transitions lasting days, e.g.\ $\sim$ 1 day in 4U 0115+63 \citep{Campana_2001}. They are observed in low-mass X-ray binaries, such as 4U 1626--67, which accrete via a persistent disk \citep{Genccali_2022}; symbiotic X-ray binaries, such as GX 1+4 \citep{Gonzalez_2012}; and high-mass X-ray binaries, such as Vela X--1, which accrete via a transient, usually wind-fed disk \citep{Serim_2023}. The secular episodes of acceleration and deceleration are correlated sometimes with X-ray outbursts and quiescence respectively, e.g.\ in A 0535+26, but not always \citep{Jain_2010,Mushtukov_2024}. The physical cause of torque reversals remains uncertain. Plausible mechanisms include a retrograde accretion disk \citep{Nelson_1997,Murray_1999,Jenke_2012}, a magnetically warped disk \citep{Matsuda_1987,Makishima_1988,VanKerkwijk_1998,Lai_1999,Beri_2014}, a quasi-Keplerian disk with an extended magnetic dynamo \citep{Habumugisha_2020}, advective or quasi-spherical accretion \citep{Yi_1997,Shakura_2012}, and versions of the magnetocentrifugal propeller effect involving magnetic obliquity and state transitions mediated by hydromagnetic instabilities at the disk-magnetosphere boundary \citep{Illarionov_1975,Fritz_2006,Perna_2006,Bozzo_2008,Dangelo_2010,Dangelo_2012,Romanova_2015}.

The standard theory of magnetocentrifugal accretion accommodates positive and negative torques even in the absence of retrograde rotation or warping in the accretion disk \citep{Ghosh_1979,Dangelo_2010}. The time derivative of the star's angular velocity $\Omega(t)$ scales as $\dot{\Omega}(t) \propto Q(t) R_{\rm{m}}(t)^{1/2} n[\omega(t)]$, where $Q(t)$ is the mass accretion rate, $R_{\rm{m}}(t)$ is the magnetospheric radius, $\omega(t)=[R_{\rm{m}}(t)/R_{\rm{c}}(t)]^{3/2}$ is the fastness, $R_{\rm{c}}(t)$ is the corotation radius, and $n(\omega)$ is a dimensionless function which takes either sign, e.g.\ $n(\omega) = 1 - \omega$. In this picture, if an X-ray pulsar undergoes a torque reversal, one must have $0 < \omega(t) < 1$ during the acceleration episode [$\dot{\Omega}(t)>0$] and $1 < \omega(t) \lesssim 1.25$ during the deceleration episode [$\dot{\Omega}(t) < 0$], i.e.\ the weak propeller regime where the star spins down but continues to pulsate \citep{Illarionov_1975,Lovelace_1999,Romanova_2005,Ustyugova_2006,Romanova_2015,Romanova_2018}. However, traditional analyses of $\Omega(t)$ timing observations cannot infer $\omega(t)$ uniquely; $R_{\rm{m}}(t)$ and hence $\omega(t)$ are degenerate with $Q(t)$ because they appear as an inseparable product in $\dot{\Omega}(t)$, and $Q(t)$ is degenerate with the radiative efficiency $\eta(t)$ in the aperiodic X-ray flux $L(t) \propto Q(t) \eta(t)$ \citep{Frank_2002,klochkov_2009,doroshenko_2010,Longair_2010,Melatos_2022,Mushtukov_2024}. Hence one cannot check observationally, with a traditional analysis, whether the magnetocentrifugal picture is self-consistent, i.e.\ whether the system occupies the regimes $0 < \omega(t) < 1$ and $1 < \omega(t) \lesssim 1.25$ during acceleration and deceleration respectively. Similar degeneracies interfere with consistency checks involving the star's magnetic moment $\mu \propto \Omega(t)^{-7/6}Q(t)^{1/2}$ \citep{Ghosh_1979}, which should remain approximately constant and equal during the acceleration and deceleration episodes. 

Recently it has been shown that one can resolve the degeneracies discussed in the previous paragraph by analyzing fluctuations in $\Omega(t)$ and $L(t)$ with a Kalman filter \citep{Melatos_2022}. The approach complements related but different X-ray pulsar parameter estimation studies employing nonoptimal $\chi^2$ estimators \citep{Takagi_2016,Yatabe_2018}, Bayesian analyses of accretion torque models \citep{karaferias_2023}, and spectral analyses of X-ray timing observations \citep{Serim_2022,Serim_2023}. Kalman filter analyses of Rossi X-ray Timing Explorer (RXTE) observations of 24 X-ray pulsars in the Small Magellanic Cloud \citep{Yang_2017} have produced independent measurements of $\mu$ and $\eta(t)$ for the 24 objects \citep{OLeary_2024a,OLeary_2024b} and resolved stable, chaotic unstable, and ordered unstable Rayleigh-Taylor accretion regimes for the first time \citep{Blinova_2016,OLeary_2024c}. In this paper, we apply the Kalman filter framework developed by \cite{Melatos_2022} to $N_{\rm{B}} = 706$ CGRO and $N_{\rm{F}} = 2634$ FGRST measurements of the pulse frequency $\nu(t) \propto \Omega(t)$ of the low-mass X-ray binary 4U 1626--67 on both sides of the 2008 torque reversal \citep{Camero_2009,Jain_2010,Camero_2012,Beri_2014,turkouglu_2017,Benli_2020,Genccali_2022,Serim_2023} and perform consistency tests of the magnetocentrifugal picture involving $\omega(t)$ and $\mu$, as discussed in the previous paragraph. In Section \ref{Sec:Observations} we introduce the post-processed CGRO and FGRST $\nu(t)$ time series analyzed in this paper. In Section \ref{Sec:MagDyn}, we explain how to formulate the Kalman filter to analyze a $\nu(t)$ time series  without $L(t)$ samples. We do this to take advantage of the $\nu(t)$-only CGRO and FGRST datasets, introduced in Section \ref{Sec:Observations}, which contain $\sim 3 \times 10^3$ samples, compared to $< 10^3$ samples per pulsar in the $\nu(t)$-and-$L(t)$ RXTE dataset \citep{Yang_2017}. In Section \ref{Sec:Torque}, we run the Kalman filter twice, once assuming a prograde accretion disk during the  deceleration and acceleration episodes, and once assuming retrograde and prograde disks during deceleration and acceleration, respectively. We infer $Q(t)$, $\dot{\Omega}(t)$, and $\omega(t)$ under both scenarios and assess them for consistency. Astrophysical implications for magnetocentrifugal accretion are canvassed briefly in Section \ref{Sec:Conclusion}.

\section{X-ray timing observations of 4U 1626--67}\label{Sec:Observations}

4U 1626--67 was discovered by the UHURU X-ray observatory \citep{Giacconi_1972}. The system contains a persistent $130$ mHz X-ray pulsar \citep{Rappaport_1977} and a low-mass ($< 0.1 M_{\odot}$), hydrogen-depleted \citep{Levine_1988,Chakrabarty_1998}, optical stellar companion, KZ TrA \citep{Mcclintock_1977,Ilovaisky_1978,Mcclintock_1980}. The system is ultracompact, with orbital period $\approx 42 \, \rm{min}$  \citep{Middleditch_1981}. Its measured optical parallactic distance is $3.5^{+2.3}_{-1.3}$ kpc \citep{Bailer_2018}; see Section 4.2 of \cite{Schulz_2019} for details about inferring the distance to 4U 1626--67.  

We focus our analysis on 4U 1626--67 for the following three practical reasons. (i) It accretes via a disk \citep{Chakrabarty_19974U}, consistent with the canonical picture of magnetocentrifugal accretion \citep{Ghosh_1977,Ghosh_1979} as well as the accretion dynamics presented in Section \ref{Sec:MagDyn} below. (ii) There are $N = N_{\rm{B}} + N_{\rm{F}} = 3340$ post-processed CGRO and FGRST $\nu(t)$ samples publicly available, the highest number of samples currently available via the CGRO and FGRST databases for persistent X-ray pulsars. (iii) Spectral analysis of BeppoSAX observations \citep{Boella_1997} attributes an absorption feature of the pulse-averaged spectrum of 4U 1626--67 to electron cyclotron transitions, yielding a direct and independent measurement of the surface magnetic field strength $B \approx 3 \times 10^{12} \, \rm{G}$ and hence magnetic dipole moment $\mu$ \citep{Orlandini_1998}. Independent measurements of $B$ and $\mu$ are not available for every high-mass X-ray binary in the Small Magellanic Cloud studied previously \citep{OLeary_2024a,OLeary_2024b,OLeary_2024c}. 

X-ray timing data from the CGRO Burst And Transient Source Experiment [BATSE; \cite{Finger_1997}] Pulsar Team\footnote{\href{https://gammaray.nsstc.nasa.gov/batse/pulsar/}{https://gammaray.nsstc.nasa.gov/batse/pulsar/}\label{FN:BATSE}} and the Fermi Gamma-Ray Burst Monitor [GBM; \cite{Malacaria_2020}] Accreting Pulsar Program\footnote{\href{https://gammaray.nsstc.nasa.gov/gbm/science/pulsars.html}{https://gammaray.nsstc.nasa.gov/gbm/science/pulsars.html}\label{FN:GBM}} contain important time-dependent information about how the pulse frequency $\nu(t)$ fluctuates stochastically in response to the magnetocentrifugal accretion torque \citep{Deeter_1989,Baykal_1993,Bildsten_1997}. The BATSE and GBM payloads have collected X-ray timing observations of tens of accretion-powered pulsars in the 20--70 keV and 10--50 keV energy range for $\approx 9$ yr and $\approx 16$ yr, respectively; see Footnotes \ref{FN:BATSE} and \ref{FN:GBM} for the names and timing properties of the X-ray pulsars monitored by the BATSE and GBM.\footnote{Additionally, post-processed BATSE and GBM data products have led to the discoveries of $\sim 10^3$ gamma-ray bursts \citep{Von_2020} and their cosmological origin \citep{Fishman_1995,Meegan_1998}, and have been instrumental in gravitational-wave multi-messenger astrophysics \citep{Connaughton_2016} as well as estimating the orbital, stellar, and timing noise properties of accretion-powered pulsars \citep{Bildsten_1997,Wilson_2018,Serim_2023}.} We refer the reader to \cite{Bildsten_1997} and \cite{Malacaria_2020} for detailed overviews of the BATSE and GBM data reduction pipelines, respectively. 

\begin{figure}
\centering{
    \includegraphics[width=\textwidth, keepaspectratio]{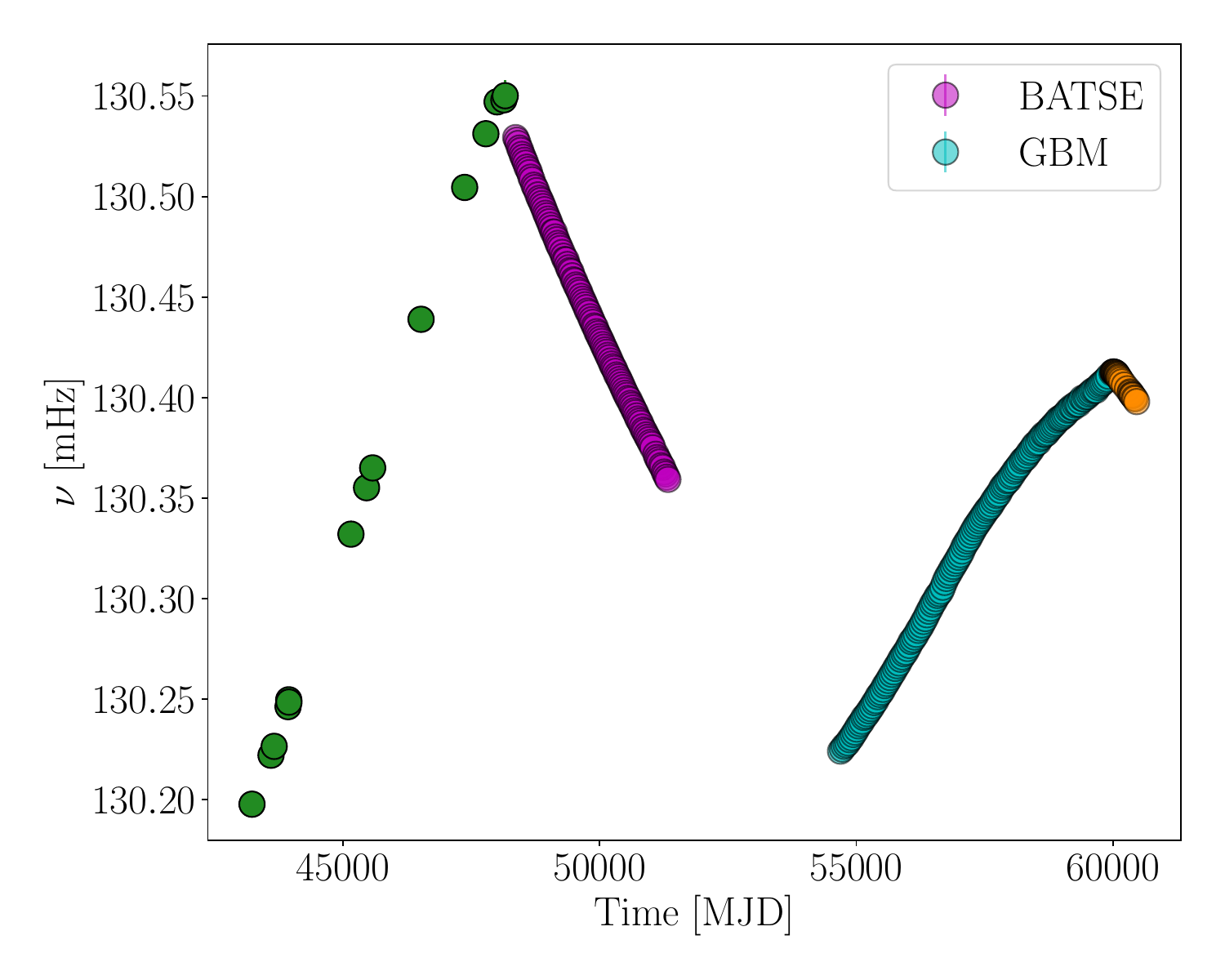}}
    \caption{Pulse frequency $\nu(t)$ observations (units: mHz) of the low-mass X-ray binary 4U 1626--67 versus time $t$ (units: MJD). The data span $\approx 47$ yr of publicly available, post-processed timing observations. Points plotted in magenta and cyan correspond to the $N_{\rm{B}} = 706$ BATSE and $N_{\rm{F}} = 2634$ GBM $\nu(t)$ samples analyzed in Section \ref{Sec:Torque}. The magenta and cyan points bracket the 2008 torque
    reversal near MJD 54500 \citep{Camero_2009}}. Green and orange points correspond respectively to 15 pre-BATSE  measurements collected by 10 X-ray observatories [see Table 1 of \cite{Chakrabarty_19974U}] and 30 GBM measurements following the most recent torque reversal around MJD 60020 \citep{Jenke_2023,Sharma_2023,Tobrej_2024}. They are not analyzed in Section \ref{Sec:Torque} and are included as a visual aid only. Error bars are plotted but are too small to be visible.
    \label{fig1}
\end{figure}

In Figure \ref{fig1} we plot the publicly available, post-processed pulse frequency $\nu(t)$ measurements of 4U 1626--67, spanning $\approx 47 \, \rm{yr}$ of timing observations. Each plotted point corresponds to a single $\nu(t)$ time sample and its associated error bars (too small to be visible; typically $\lesssim 1 \, \rm{\mu Hz}$). Points plotted in green (at left) and orange (at right) correspond to 15 historical $\nu(t)$ samples collected by 10 X-ray timing instruments [details of which can be found in Table 1 of \cite{Chakrabarty_19974U}] and 30 GBM $\nu(t)$ samples following the most recent torque reversal near MJD 60020 \citep{Jenke_2023,Sharma_2023,Tobrej_2024}, respectively. We do not consider the foregoing 45 $\nu(t)$ samples in the analysis in Section \ref{Sec:Torque}; they are plotted purely as a visual aid for the convenience of the reader. Similarly, we do not analyze the $\sim 10^2$ $\nu(t)$ samples covering the 2008 torque reversal between MJD 53300 and MJD 54900 in Figure 3 of \cite{Camero_2009}, because they are not publicly available in post-processed form. They are mentioned here to support future Kalman filter studies.  Here we analyze the $N_{\rm{B}} = 706$ BATSE (magenta points) and $N_{\rm{F}} = 2634$  GBM (cyan points) publicly available $\nu(t)$ samples associated with the $\approx 8.2 \, \rm{yr}$ deceleration and $\approx 14.5 \, \rm{yr}$ acceleration episodes bracketing the 2008 torque reversal near MJD 54500 \citep{Camero_2009,Jain_2010,Camero_2012,Beri_2014,turkouglu_2017,Benli_2020,Genccali_2022,Serim_2023}. The $N = N_{\rm B} + N_{\rm F} = 3340$ points are spaced unequally, with minimum separation $\approx 1.3 \, \rm{days}$ and maximum separation $\approx 28 \, \rm{days}$.

\section{Magnetocentrifugal accretion dynamics}\label{Sec:MagDyn}

In this section, we reformulate the canonical theory of magnetocentrifugal accretion as an equivalent state-space model, which can be analyzed readily with a Kalman filter. Specifically, we supplement the canonical magnetocentrifugal accretion torque law \citep{Ghosh_1977,Ghosh_1979} with a phenomenological, idealized model of the stochastic fluctuations in the mass accretion rate and Maxwell stress at the disk-magnetosphere boundary \citep{Melatos_2022}, and relate the system variables to the observed time series $\nu(t)$. The framework is a special case of the one developed by \cite{Melatos_2022} to analyze $\nu(t)$ and $L(t)$ time series simultaneously, except that we generalize the torque law to accommodate retrograde and prograde accretion disk configurations \citep{Nelson_1997,Psaltis_1999}. 

We express the standard picture of magnetocentrifugal accretion in terms of three time-dependent magnetospheric variables. The angular velocity of the star $\Omega(t)$ (units: $\rm{rad \, s^{-1}}$) is measured directly, viz.\
\begin{equation}\label{eq:nu}
    \nu(t) = \Omega(t)/(2 \pi) + N_{\nu}(t).
\end{equation}
The additive measurment noise term in Equation (\ref{eq:nu}) is assumed to be Gaussian and white, with $\langle N_{\nu}(t_n) \rangle = 0$, and $\langle N_{\nu}(t_n) N_{\nu}(t_{n'})  \rangle = \Sigma^2_{\nu\nu} \delta_{n, n'}$, where $\delta_{n, n'}$ denotes the Kronecker delta. The other two magnetospheric variables are hidden and hence cannot be measured directly. The mass accretion rate is denoted by $Q(t)$ (units: ${\rm{g \, s^{-1}}}$). The scalar Maxwell stress at the disk-magnetosphere boundary is denoted by $S(t)$ (units: $\rm{g \, cm^{-1} \, s^{-2}}$).

The traditional picture of magnetocentrifugal accretion \citep{Ghosh_1977,Ghosh_1979} assumes a prograde disk. It gives rise to an accretion phase with $R_{\rm{m}}(t) < R_{\rm{c}}(t)$ and $\dot{\Omega}(t) > 0$, and a propeller phase with $R_{\rm{m}}(t) > R_{\rm{c}}(t)$ and $\dot{\Omega} < 0$ \citep{Illarionov_1975,Lovelace_1999, Romanova_2003a,Romanova_2004a,Ustyugova_2006, Romanova_2018}. In the accretion phase, material at the disk-magnetosphere boundary orbits faster than the star, penetrates the magnetosphere along open field lines, and lands at antipodal hotspots near the two magnetic poles. Three-dimensional magnetohydrodynamic simulations indicate that multiple, transient, complicated funnels as well as wandering hot spots are also possible \citep{Romanova_2008,Kulkarni_2008,Romanova_2015,Blinova_2016}. In the propeller phase, material at the disk-magnetosphere boundary orbits slower than the star, and some of it is ejected centrifugally to form an outflow \citep{Matt_2005, Romanova_2005,Lovelace_2014}. Simulations indicate that some material also penetrates the magnetosphere and reaches the surface for $1\leq \omega(t) \lesssim 1.25$ \citep{Spruit_1993,Dangelo_2010,Lii_2014,Papitto_2015}. In this paper, we retain the features above and simply generalize the canonical hydromagnetic torque law \citep{Ghosh_1977,Ghosh_1979} with a prefactor $\epsilon(t)$, viz.\
\begin{equation} \label{eq:torque}
    I \frac{d \Omega}{dt} = \epsilon(t) Q(t) [GM R_{\rm{m}}(t)]^{1/2} \, n (\omega), 
\end{equation}
to allow for either a prograde [$\epsilon(t) = 1$] or retrograde [$\epsilon(t) = -1$] disk \citep{Nelson_1997,Psaltis_1999}. In Equation (\ref{eq:torque}) we take $\omega(t) = [R_{\rm{m}}(t)/R_{\rm{c}}(t)]^{3/2}$, $R_{\rm m} (t) = (2 \pi^{2/5})^{-1} (GM)^{1/5} Q(t)^{2/5} S(t)^{-2/5}$, $R_{\rm c}(t) = (GM)^{1/3}\Omega(t)^{-2/3}$, and $n(\omega) = 1 - \omega(t)$ \citep{Melatos_2022,OLeary_2024a,OLeary_2024b}. We emphasize that Equation (\ref{eq:torque}) is highly idealized, and more realistic alternatives exist, e.g.\ which lead to phenomena like episodic accretion and disk trapping \citep{Dangelo_2010,Dangelo_2012,Dangelo_2017}. Such phenomena ought to be included, when datasets become available that are substantially larger than those analyzed in this paper. 

The disk-magnetosphere interaction is complicated geometrically, e.g.\ due to disk warping and precession induced by the misalignment of the rotation and magnetic axes \citep{Foucart_2011}. It involves complicated, stochastic, hydromagnetic processes, e.g.\ self-healing, magnetic, Rayleigh-Taylor instabilities at the disk-magnetosphere boundary \citep{Arons_1976,Anzer_1980,Blinova_2016,OLeary_2024c}. To make progress, we adopt a phenomenological, idealized, Ornstein-Uhlenbeck model \citep{Gardiner_1985} of the stochastic processes at the disk-magnetosphere boundary, and postulate mean-reverting dynamics in $Q(t)$ and $S(t)$. That is, we assume that $Q(t)$ and $S(t)$ satisfy the Langevin equations
\begin{eqnarray}
    \frac{dQ}{dt} &= -\gamma_Q [Q(t) - \bar{Q}] + \xi_{Q}(t), \label{eq:Lang_Q}\\
    \frac{dS}{dt} &= -\gamma_S [S(t) - \bar{S}] + \xi_{S}(t) \label{eq:Lang_S},
\end{eqnarray}
with $\langle \xi_{A}(t) \rangle = 0$ and $\langle \xi_{A}(t) \xi_{B}(t') \rangle = \sigma_{AB}^2 \delta_{AB} \delta (t - t')$, where the Kronecker and Dirac delta functions are denoted by $\delta_{AB}$ and $\delta(t - t')$, respectively. Here $\gamma_A^{-1}$ denotes the characteristic timescale of mean-reversion about the asymptotic, ensemble-averaged value $\langle A(t) \rangle \approx \bar{A}$, and $\sigma_{AA}$ parameterizes the strength of the white noise drivers, leading to rms fluctuations $\sim \gamma_{A}^{-1/2} \sigma_{AA}$ in $A(t)$ for $A$ $\in \{Q, S \}$. Although idealized, Equations (\ref{eq:Lang_Q}) and (\ref{eq:Lang_S}) have proved successful in analyzing RXTE data from X-ray pulsars in the Small Magellanic Cloud to infer their magnetic moments \citep{OLeary_2024a,OLeary_2024b}, radiative efficiency \citep{OLeary_2024a,OLeary_2024b}, and Rayleigh-Taylor accretion regimes \citep{OLeary_2024c}. 

Equations (\ref{eq:torque})--(\ref{eq:Lang_S}) neglect important aspects of realistic accretion physics, including the tensorial nature of the Maxwell stress, the nonzero thickness of the disk-magnetosphere boundary, the rapid onset of the propeller transition and disk trapping in the weak propeller regime \citep{Dangelo_2010,Dangelo_2012,Dangelo_2015,Dangelo_2017}, as well as outflows \citep{Matt_2005,Matt_2008} and disk warping and precession induced by the misalignment of the magnetic and rotation axes \citep{Foucart_2011,Romanova_2021}. These idealizations and others are discussed in detail in Section 2.4 and Appendix C of \cite{Melatos_2022} and Section 2.3 of \cite{OLeary_2024b}  and are not discussed further here. 

\section{Kalman filter analysis of the 2008 torque transition}\label{Sec:Torque}
The measurement process described by Equation (\ref{eq:nu}) and the dynamical process described by Equations (\ref{eq:torque})--(\ref{eq:Lang_S}) take a mathematical form, which matches exactly the requirements of a Kalman filter \citep{Kalman1960,Gelb_1974,Julier_1997}. That is, given a measured time series $\bm{Y} = [\nu(t_n)]$ for $1 \leq n \leq N$, one employs a Kalman filter to estimate the most probable state sequence $\Omega(t_n)$, $Q(t_n)$, and $S(t_n)$ consistent with Equations (\ref{eq:nu})--(\ref{eq:Lang_S}). As Equation (\ref{eq:torque}) is nonlinear, we employ an unscented Kalman filter \citep{Julier_1997,Wan_2000,Wan_2001}. We couple the unscented Kalman filter to a nested sampler \citep{Skilling_2004,Skilling_2006,Ashton_2022} to infer the most probable values of the static, model parameters $\bm{\Theta}$. Specifically, we maximize the Kalman filter likelihood to infer $\bar{Q}$, $\bar{S}$, and hence $\mu$, by applying (for example) Equation (14) of \cite{OLeary_2024b}. Detailed discussions of the Kalman filter analysis are given in Section 3.2 and Appendix D in \cite{Melatos_2022}, Section 4 and Appendix A in \cite{OLeary_2024a}, and Section 3 and Appendix A in \cite{OLeary_2024b}. They are reviewed in abridged form in Appendix \ref{App:KF_PE} here. 

In this section, we run the Kalman filter twice and perform consistency checks between two magnetocentrifugal accretion models involving $Q(t)$, $\dot{\Omega}(t)$, and $\omega(t)$. The first model, $\mathcal{M}_{\rm{PP}}$, assumes a prograde disk [$\epsilon(t) = 1$] during both the deceleration (magenta points, Figure \ref{fig1}) and acceleration (cyan points, Figure \ref{fig1}) episodes. The second model, $\mathcal{M}_{\rm{RP}}$, assumes a retrograde disk [$\epsilon(t) = -1$] during deceleration (magenta points, Figure \ref{fig1}) and a prograde disk [$\epsilon(t) = 1$] during acceleration (cyan points, Figure \ref{fig1}). In Section \ref{SubSec:ModComparison}, we perform Bayesian model selection by comparing the $\mathcal{M}_{\rm{PP}}$ and $\mathcal{M}_{\rm{RP}}$ Bayes factor and likelihood ratio returned by the nested sampler and Kalman filter, respectively. In Section \ref{SubSec:MagParams}, we assess how the inferred magnetospheric parameter $\bar{Q}$ differs during the deceleration and acceleration episodes under the $\mathcal{M}_{\rm{PP}}$ and $\mathcal{M}_{\rm{RP}}$ hypotheses. {We compare the time-resolved mass accretion rate $Q(t)$ with the aperiodic X-ray flux $F_{X}(t)$ measured by \cite{Chakrabarty_19974U}, \cite{Orlandini_1998}, and \cite{Krauss_2007} in Section \ref{SubSec:QVsFL}. We then compare the magnetocentrifugal accretion torque $\propto \dot{\Omega}(t)$ with previous X-ray timing and spectral analyses \citep{Camero_2009,Camero_2012} in Section \ref{SubSec:TROm}. In Section \ref{SubSec:HiddenState}, we measure the fastness histories $\omega(t_1),\hdots, \omega(t_{N_{\rm{B}}})$ and $\omega(t_{N_{\rm B}+1}),\hdots, \omega(t_{N_{\rm{B}}+N_{\rm{F}}})$ and resolve the Rayleigh-Taylor accretion regimes \citep{Blinova_2016} associated with $\mathcal{M}_{\rm{PP}}$ and $\mathcal{M}_{\rm{RP}}$.  

\subsection{Model selection: Bayes factor and Kalman likelihood ratio}\label{SubSec:ModComparison}

We compare the evidence in favor of the retrograde-prograde model ${\cal M}_{\rm RP}$ versus the prograde-only model ${\cal M}_{\rm PP}$ in two related ways, discussed in Appendix \ref{SubSec:InferenceOutputs}. Firstly, we calculate the log Bayes factor $\log_{10} \mathcal{B} = \log_{10} \mathcal{Z}(\bm{Y}_{\rm{B}}, \bm{Y}_{\rm{F}}| \mathcal{M}_{\rm{RP}})  - \log_{10} \mathcal{Z}(\bm{Y}_{\rm{B}}, \bm{Y}_{\rm{F}}| \mathcal{M}_{\rm{PP}})$, where $\bm{Y}_{\rm B}$ and $\bm{Y}_{\rm F}$ denote the BATSE and FGRST data respectively.\footnote{The marginal likelihood, i.e.\ the Bayesian evidence $\mathcal{Z}$, is defined according to 
\begin{equation}
\mathcal{Z} = \int d \bm{\Theta} \, p(\bm{Y}|\bm{\Theta}, \mathcal{M}) \, p(\bm{\Theta}).
\end{equation}
It is approximated in practice, e.g.\ via Equation (16) of \cite{Speagle_2020}, when using the \texttt{dynesty} nested sampler.} The result is $\log_{10}{\cal B} = 0.44$. Secondly, we calculate the logarithm of the maximum a posteriori (MAP) Kalman likelihood, $\log_{10} \Lambda = \log_{10} p(\bm{Y}_{\rm{B}},\bm{Y}_{\rm{F}}|\bm{\Theta}_{\rm{RP, MAP}}, \mathcal{M}_{\rm{RP}}) - \log_{10} p(\bm{Y}_{\rm{B}},\bm{Y}_{\rm{F}}|\bm{\Theta}_{\rm{PP, MAP}}, \mathcal{M}_{\rm{PP}})$, where ${\bf \Theta}_{\rm RP,MAP}$ and ${\bf \Theta}_{\rm PP,MAP}$ denote the MAP values of the static parameters ${\bf\Theta}$ inferred for ${\cal M}_{\rm RP}$ and ${\cal M}_{\rm PP}$, respectively. The result is $\log_{10}\Lambda = 2.5$. Both measures imply a significant albeit moderate preference for ${\cal M}_{\rm RP}$ over ${\cal M}_{\rm PP}$. We remind the reader that $\log_{10} {\cal B} \gtrsim 0.5$ is arbitrarily yet widely regarded as ``substantial evidence'' for ${\cal M}_{\rm RP}$ over ${\cal M}_{\rm PP}$ on the standard Jeffreys scale \citep{Kass_1995,Jeffreys_1998}.

\subsection{Magnetospheric accretion parameters}\label{SubSec:MagParams}

Are the MAP values of the ensemble-averaged mass accretion rate $\bar{Q}$ implied by ${\cal M}_{\rm RP}$ and ${\cal M}_{\rm PP}$ equally plausible astrophysically, or does ${\cal M}_{\rm RP}$ do somewhat better than ${\cal M}_{\rm PP}$ in this respect too? To test this we display the one-dimensional marginalized posterior distribution of $\bar{Q}$ as a histogram in Figure \ref{fig2}. Histograms plotted in gray and red correspond to $\mathcal{M}_{\rm PP}$ and $\mathcal{M}_{\rm RP}$, respectively. The left and right panels summarize the one-dimensional $\bar{Q}$ posterior distributions associated with BATSE deceleration data  (cyan points, Figure \ref{fig1}) and FGRST acceleration data (magenta points, Figure \ref{fig1})  respectively. In each panel, the horizontal and vertical axes are plotted on $\log_{10}$ and linear scales, respectively. The steps required to generate the one-dimensional $\bar{Q}$ posterior are discussed in Section 4.3 of \cite{OLeary_2024a} for the sake of reproducibility. As well as the one-dimensional, marginalized posterior distribution of $\bar{Q}$, the Kalman filter in Appendix \ref{App:KF_PE} also outputs the one-dimensional, marginalized posterior distribution of the scalar Maxwell stress at the disk-magnetosphere boundary $\bar{S}$, viz.\ Equations (\ref{Eq:ScalingQS1}) and (\ref{Eq:ScalingQS2}), and hence the magnetic dipole moment $\mu$, viz. Equation (\ref{Eq:CnstMu}). Although we do not discuss the posterior distribution of $\mu$ in the main text of the present paper (for reasons discussed in Appendix \ref{App:SysUncertainties}), we present some preliminary results about $\mu$ in Appendix \ref{App:misalignment} for aligned and misaligned rotators to support future Kalman filter analyses of accretion-powered pulsars in the Galaxy and elsewhere.

The results in Figure \ref{fig2} have three key features. First, during deceleration (left panel), the one-dimensional posterior distribution of $\bar{Q}$ associated with $\mathcal{M}_{\rm{RP}}$ (red histogram) is unimodal with a well-defined peak. The corresponding one-dimensional posterior associated with $\mathcal{M}_{\rm{PP}}$ (gray histogram)  is approximately uniform and hence weakly informative. Second, during acceleration (right panel), the one-dimensional posterior distributions of $\bar{Q}$ associated with $\mathcal{M}_{\rm{RP}}$ (red histogram) and $\mathcal{M}_{\rm{PP}}$ (gray histogram) are both unimodal with well-defined peaks. Third, the dispersion associated with the one-dimensional $\bar{Q}$ posterior of $\mathcal{M}_{\rm{RP}}$ is modest. For example, the full width half maximum (FWHM) equals 0.040 dex and 0.18 dex for $\bar{Q}$ during deceleration and acceleration, respectively. In contrast, for $\mathcal{M}_{\rm{PP}}$, we find that the FWHM of $\bar{Q}$ during deceleration equals 0.79 dex, compared to 0.17 dex during acceleration. Taken together, the three features above are consistent with ${\cal M}_{\rm RP}$ explaining the data better than ${\cal M}_{\rm PP}$. However, it is vital to emphasize that the results in Figure \ref{fig2} do not prove by themselves, that one should prefer ${\cal M}_{\rm RP}$ over ${\cal M}_{\rm PP}$; they are indicative only.

It is unclear astrophysically, whether a torque transition should be accompanied by an abrupt shift in $Q(t)$ and hence $\bar{Q}$ or not \citep{Nelson_1997,Yi_1997,Yi_1999,Dai_2006}. One cannot test this hypothesis for $\mathcal{M}_{\rm PP}$ (gray histograms; left and right panels of Figure \ref{fig2}), because the error bars are substantial. We infer $\bar{Q} = 0.44_{-0.24}^{+0.63} \times 10^{16} \, \rm{g \, s^{-1}}$ during deceleration and $\bar{Q} = 0.74^{+0.11}_{-0.14} \times 10^{16} \, \rm{g \, s^{-1}}$ during acceleration, so it is challenging to test for a shift in the peaks of the $\bar{Q}$ posteriors in a statistically meaningful way. The central values in the foregoing estimates correspond to the posterior median and the uncertainty is quantified using a 68\% credible interval. 
 In contrast, for $\mathcal{M}_{\rm{RP}}$ (red histograms; left and right panels of Figure \ref{fig2}),  the peaks of the one-dimensional $\bar{Q}$ posteriors shift by $\approx 0.34$ dex on either side of the torque reversal, with \ $\bar{Q} = 3.0^{+0.13}_{-0.11} \times 10^{16} \, \rm{g \, s^{-1}}$ during deceleration assuming a retrograde disk [$\epsilon(t)=-1$] and $\bar{Q} = 1.5^{+0.28}_{-0.21} \times 10^{16} \, \rm{g \, s^{-1}}$ during acceleration assuming a prograde disk [$\epsilon(t) = 1$]. The foregoing $\bar{Q}$ estimates associated with $\mathcal{M}_{\rm{RP}}$ are in accord with those deduced observationally \citep{Chakrabarty_19974U,Chakrabarty_1998,Sharma_2023,Tobrej_2024} or otherwise \citep{daiAnt_2017,Schulz_2019,Benli_2020}, which satisfy $1.0 \lesssim \bar{Q}/(10^{16} \, \rm{g \, s^{-1}}) \lesssim 8.0$. Moreover,  the relatively smooth shift by $\approx 0.34 \, \rm{dex}$ between $\bar{Q}_{\rm B, RP}$ and $\bar{Q}_{\rm F, RP}$ is broadly consistent with previous spectral and X-ray timing analyses of BATSE, Swift/BAT, and Fermi/GBM measurements of 4U 1626$-$67 during the 1990 \citep{Wilson_1993,Bildsten_1994} and 2008 \citep{Camero_2009} torque reversals near MJD 48000 and MJD 54500, respectively. Previous analyses found that $\dot{\Omega}(t) \propto Q(t) \, R_{\rm m}(t)^{1/2} \, n[\omega(t)]$ shifted by $\lesssim 20\%$ during the 1990 and 2008 torque reversals, a relatively smooth transition which offers further, indirect evidence that $\mathcal{M}_{\rm RP}$ explains the data better than $\mathcal{M}_{\rm PP}$ (see Section \ref{SubSec:TROm} for details). 

We draw the reader's attention to the following important point. Although the new results in Figure \ref{fig2}, and the $\mathcal{B}$ and $\Lambda$ calculations in Section \ref{SubSec:ModComparison}, imply that $\mathcal{M}_{\rm RP}$ describes the data $\bm{Y}$ better than $\mathcal{M}_{\rm PP}$, one cannot definitively rule out the possibility that the transition from deceleration to acceleration is accompanied by an abrupt shift in $Q(t)$. It is possible to imagine reasonable physical scenarios, where $Q(t)$ makes an abrupt transition at the torque reversal, if (for example) there is a sudden transformation in the geometry of the disk-magnetosphere boundary, driven by a global hydromagnetic instability \citep{Li_1998, VanKerkwijk_1998, Lai_1999}. Under such circumstances, it would be a coincidence, that $\dot{\Omega}(t)$ (Section \ref{SubSec:TROm}) and $\omega(t)$ (Section \ref{SubSec:HiddenState}) transition relatively smoothly, while $Q(t)$ transitions abruptly.

\begin{figure}
\centering{
    \includegraphics[width=\textwidth, keepaspectratio]{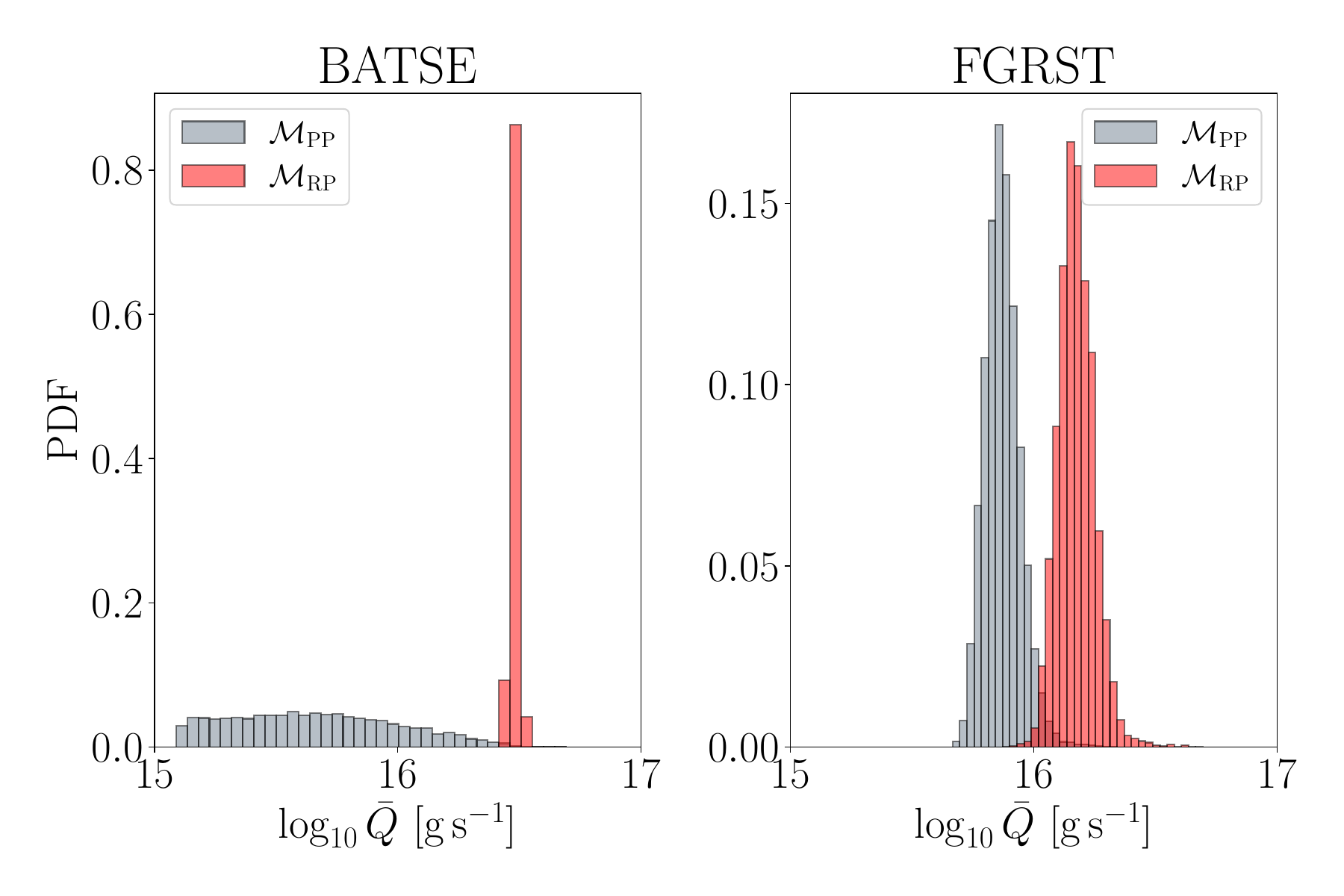}}
    \caption{One-dimensional marginalized posterior distribution of the mass accretion rate $\bar{Q}$ (units: $\rm{g \, s^{-1}})$ of the persistent X-ray pulsar 4U 1626--67 for $\mathcal{M}_{\rm PP}$ (gray histograms) and $\mathcal{M}_{\rm RP}$ (red histograms). The left and right panels summarize the $\bar{Q}$ posterior distributions during the BATSE deceleration and the FGRST acceleration episodes, respectively. The vertical and horizontal axes are on linear and $\log_{10}$ scales, respectively. The acronym PDF stands for probability density function.}
    \label{fig2}
\end{figure}

\subsection{Time-resolved mass accretion rate $Q(t)$}\label{SubSec:QVsFL}

One advantage of a Kalman filter analysis is that it returns time-resolved estimates of the state variables, e.g.\ the mass accretion rate $Q(t)$, throughout the observation span. One can therefore compare the performance of the models $\mathcal{ M}_{\rm PP}$ and $\mathcal{M}_{\rm RP}$ not only on the basis of what they predict about the static parameters, such as $\bar{Q}$ in Figure \ref{fig2}, but also according to whether the time-resolved dynamics they predict, e.g.\ for $Q(t)$, are plausible physically. High-time resolution measurements of the aperiodic X-ray flux $F_X(t)$ [or aperiodic X-ray luminosity $L(t)$] of accretion-powered pulsars shed light on the physics at the disk-magnetosphere boundary \citep{Melatos_2022}.  Within the canonical magnetocentrifugal paradigm \citep{Ghosh_1977,Ghosh_1978,Ghosh_1979}, the aperiodic X-ray flux $F_{X}(t)$ scales as $F_{X}(t) \propto L(t) \propto Q(t) \, \eta(t)$.\footnote{In many non-Kalman analyses, it is standard practice to assume $\eta(t) = \bar{\eta} = \rm{constant}$. Hence, the mass accretion rate $Q(t)$ serves as a proxy for the X-ray flux, viz. $F_{X}(t) \propto L(t) \propto Q(t)$ \citep{Ho_2014,Klus_2014,Mukherjee_2015}.} Traditional X-ray timing and spectral techniques, e.g.\ quasiperiodic oscillation analyses \citep{Alpar_1985,VanDerKlis_1985,Shaham_1987, Sharma_2025,Zhou_2025}, return time-averaged point estimates of important magnetocentrifugal quantities, e.g.\ $Q$ \citep{Benli_2020,Tobrej_2024}, $\mu$ \citep{Camero_2012}, and $\dot{\Omega}$ \citep{Camero_2009}, and hence it is challenging to test the foregoing scaling in a time-resolved sense. In contrast, the Kalman filter in Appendix \ref{App:KF_PE} outputs time-resolved estimates of the state vector $\bm{X}_{n} = [\Omega(t_n), Q(t_n), S(t_n)]$ at time $t_n$. That is, for every $\nu(t_n)$ measurement in Figure \ref{fig1}, the Kalman filter returns $\bm{X}_{n} = [\Omega(t_n), Q(t_n), S(t_n)]$ for $1 \leq n \leq N = N_{\rm{B}} + N_{\rm F} = 3340$, viz. Equation (A17) of \cite{OLeary_2024b}.  We refer the reader to Section 3 and Appendix D of \cite{Melatos_2022} for a summary of the linear Kalman filter, and to Appendices A and B of  \cite{OLeary_2024b} and \cite{kimpson_2025} for summaries of nonlinear (extended and unscented) Kalman filters \citep{Julier_1997,Wan_2001,Zarchan_2005}.

In Figure \ref{fig3} we plot $Q(t_{n'})/\bar{Q}$ versus $t_{n'}$ assuming prograde [$\mathcal{M}_{\rm PP}$; $\epsilon(t_{n'}) = 1$; magenta points; top panel] and retrograde disks [$\mathcal{M}_{\rm RP}$; $\epsilon(t_{n'}) = -1$; magenta points; middle panel] during deceleration (BATSE observations; magenta points, Figure \ref{fig1}) with $1 \leq n' \leq  N_{\rm B}$. The time series $Q(t_{n'})$ in the top and middle panels of Figure \ref{fig3} are normalized by $\bar{Q}$, the mode of the one-dimensional, marginalized posterior $\bar{Q}$, visible as a gray ($\mathcal{M}_{\rm PP}$) or red ($\mathcal{M}_{\rm RP}$) histogram in the left panel of Figure \ref{fig2}. In the bottom panel of Figure \ref{fig3}, we plot $F_{X}(t_{m'})/F_{\rm HEAO}$ versus $t_{m'}$, where $m'$ coincides with the seven flux measurements reported in Table \ref{tab1}.\footnote{The flux measurements $F_X(t_{m'})$ plotted in the bottom panel of Figure \ref{fig3} are reported in Table 2 of \cite{Chakrabarty_19974U} (black stars), Section 3.1 of \cite{Orlandini_1998} (black squares), and Table 2 of \cite{Krauss_2007} (black diamonds). The $1\sigma$ error bars are reported in \cite{Chakrabarty_19974U} only.} \footnote{We adopt the broken power-law spectral model of \cite{Pravdo_1979} with photon index $\Gamma = 1.6$ and assume a Galactic hydrogen column density $n_{\rm H} = 9.6 \times 10^{20} \, \rm{cm^{-2}}$ \citep{Sharma_2025}.} We normalize $F_{X}(t_{m'})$ by $F_{\rm HEAO} = 26 \times 10^{-10} \, \rm{erg \, cm^{-2} \, s^{-1}}$, the aperiodic X-ray flux measured by HEAO 1 near MJD 43596 \citep{Pravdo_1979,Chakrabarty_19974U}. The X-ray flux measurements are measured in four different energy ranges (Table \ref{tab1}, third column). We convert $F_X(t_{m'})$ to a common energy range 0.7--60 keV \citep{Pravdo_1979} using the High Energy Astrophysics Science Archive Research Center WebPIMMs tool.\footnote{\href{https://heasarc.gsfc.nasa.gov/cgi-bin/Tools/w3pimms/w3pimms.pl}{https://heasarc.gsfc.nasa.gov/cgi-bin/Tools/w3pimms/w3pimms.pl}}

The time-resolved results in Figure \ref{fig3} have three key features. First, $Q(t_{n'})$ associated with $\mathcal{M}_{\rm PP}$ remains almost constant during deceleration, fluctuating within $\approx 1\%$ of $\bar{Q}$ for the total observation time, i.e.\ $Q(t_{n'}) \approx \bar{Q}$. That is, the results in the top panel of Figure \ref{fig3} suggest that the radiative efficiency $\eta(t)$ decreases secularly during deceleration to match $F_X(t)$ in the bottom panel, viz. $F_X(t) \propto Q(t) \, \eta(t)$. Second, $Q(t_{n'})$ associated with $\mathcal{M}_{\rm RP}$ exhibits a trend similar to $F_{X}(t)$. That is, $Q(t_{n'})$ and $F_{X}(t_{m'})$ decrease steadily between MJD 48500 and MJD 50000 and fluctuate around $Q(t_{n'})/\bar{Q} \approx 0.7$ and $F_X(t_{m'})/F_{\rm HEAO} \approx 0.2$ after MJD 50000. Third, we infer a mean-reversion timescale $\gamma_{Q,{\rm PP}} = 2.7^{+0.5}_{-1.8} \times 10^{-6} \, \rm{s^{-1}}$ and rms noise amplitude $\sigma_{QQ,{\rm PP}} = 1.4^{+1.1}_{-1.2} \times 10^{11} \, {\rm g \, s^{3/2}}$ assuming a prograde disk during deceleration and $\gamma_{Q,{\rm RP}} = 1.1^{+0.23}_{-0.10} \times 10^{-8} \, \rm{s^{-1}}$ and    $\sigma_{QQ, {\rm RP}} = 4.3^{+1.0}_{-0.8} \times 10^{11} \, \rm{g \, s^{3/2}}$ assuming a retrograde disk. It is interesting astrophysically that $\sigma_{QQ,{\rm PP}}$ and $\sigma_{QQ,{\rm RP}}$ are comparable in magnitude, yet mean reversion occurs over $\sim \rm{days}$ for $\mathcal{M}_{\rm PP}$ and $\sim \rm{years}$ for $\mathcal{M}_{\rm RP}$.
\begin{figure}
\centering{
    \includegraphics[width=\textwidth, keepaspectratio]{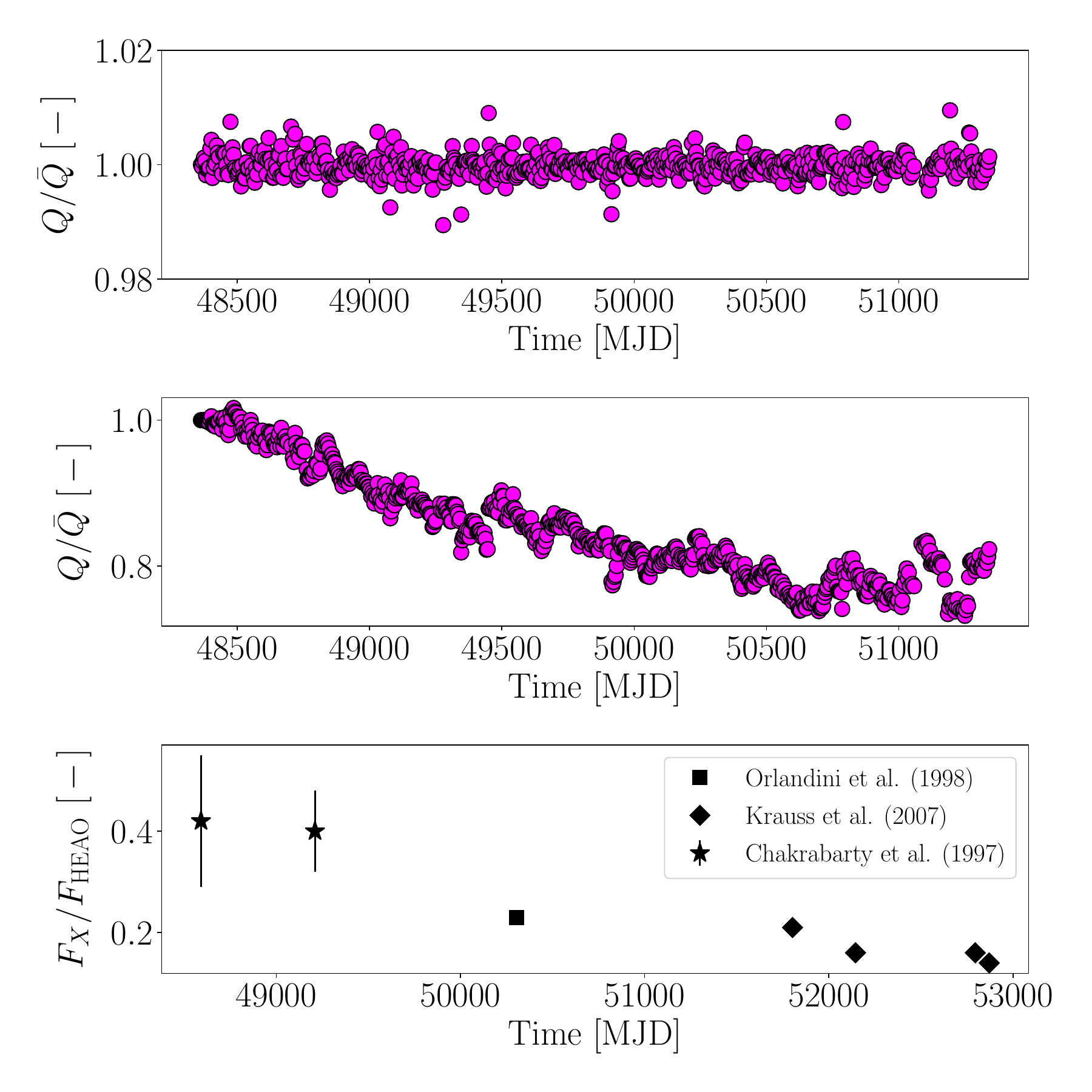}}
    \caption{Mass accretion rate $Q(t)$ (units: $\rm{g \, s^{-1}}$) and aperiodic X-ray flux $F_X(t)$ (units: $\rm{erg \, cm^{-2} \, s^{-1}}$) fluctuation histories of 4U 1626$-$67. (Top and middle panels.) Time-resolved mass accretion rate $Q(t_{n'})/\bar{Q}$ (units: dimensionless; magenta points) versus $t_{n'}$ (units: MJD) assuming prograde [$\mathcal{M}_{\rm PP}$; $\epsilon(t_{n'}) = 1$; top panel] and retrograde [$\mathcal{M}_{\rm RP}$; $\epsilon(t_{n'}) = -1$; middle panel] disks during deceleration with $1 \leq n' \leq N_{\rm B}$. The time-resolved $Q(t_{n'})$ histories are normalized by the mode of the one-dimensional marginalized posteriors  $\bar{Q}$, visible as gray ($\mathcal{M}_{\rm PP}$) or red ($\mathcal{M}_{\rm RP}$) histograms in the left panel of Figure \ref{fig2}. (Bottom panel.) Measured aperiodic X-ray flux $F_X(t_{m'})/F_{\rm{HEAO}}$ (units: dimensionless; energy range: 0.7--60 keV) versus $t_{m'}$ (units: MJD) for the seven MJD values in Table \ref{tab1}. Measurements plotted as black stars, squares, and diamonds, are reported in Table 2 of \cite{Chakrabarty_19974U}, Section 3.1 of \cite{Orlandini_1998}, and Table 2 of \cite{Krauss_2007}, respectively, as well as the fourth and fifth columns of Table \ref{tab1}. The 1$\sigma$ error bars are reported for the Ginga and ASCA measurements (black stars) only \citep{Chakrabarty_19974U}.  The $F_X(t_{m'})$ measurements are normalized by $F_{\rm HEAO} = 26 \times 10^{-10} \, \rm{erg \, cm^{-2} \, s^{-1}}$, the aperiodic X-ray flux measured by HEAO 1 in the 0.7--60 keV energy range \citep{Chakrabarty_19974U}, consistent with previous analyses of flux measurements associated with 4U 1626$-$67 \citep{Chakrabarty_19974U,Camero_2009,Camero_2012}. The horizontal and vertical axes are plotted on linear scales.}
    \label{fig3}
\end{figure}

\begin{table}
\begin{center}
\begin{tabular}{ccccc}
\hline
MJD & Instrument & Bandpass [keV] & $F_X(t)$ [$10^{-10}\, \rm{erg \, cm^{-2} \, s^{-1}}$] & $F_X(t)/F_{\rm HEAO}$ \\ \hline
48591 & Ginga/ASM & 2--20 & 6.7 $\pm$ 0.6 & 0.42 $\pm$ 0.13  \\
49210 & ASCA & 0.5--10 & 2.9 $\pm$ 0.2 & 0.40 $\pm$ 0.08  \\
50304 & BeppoSAX & 2--10 & 1.7 & 0.23 \\
51803 & Chandra & 0.3--10 & 2.2 & 0.21  \\
52145 & XMM-Newton & 0.3--10 & 1.7 & 0.16  \\
52795 & Chandra &  0.3--10 & 1.7 & 0.16 \\
52871 & XMM-Newton & 0.3--10 & 1.5 & 0.14 \\ \hline
\end{tabular}
\end{center}
\caption{Aperiodic X-ray flux $F_X(t)$ of 4U 1626$-$67 during deceleration between the 1990 and 2008 torque reversals \citep{Camero_2009,Camero_2012}. We report the MJD in the first column,  the X-ray timing instrument in the second column, the observing bandpass (units: keV) in the third column, the measured flux $F_X(t)$ (units: $\rm{erg \, cm^{-2} \, s^{-1}}$) in the fourth column, and the converted 0.7--60 keV dimensionless flux $F_X(t)/F_{\rm{HEAO}}$ in the fifth column. The flux measurements are normalized by $F_{\rm HEAO} = 26 \times 10^{-10} \, \rm{erg \, cm^{-2}\, s^{-1}}$, the aperiodic X-ray flux measured by HEAO 1 near MJD 43596 \citep{Pravdo_1979,Chakrabarty_19974U}. The foregoing normalization is arbitrary, and is adopted for consistency with previous analyses of flux measurements of 4U 1626$-$67 \citep{Chakrabarty_19974U,Camero_2009,Camero_2012}}.
\label{tab1}
\end{table}

\subsection{Time-resolved magnetocentrifugal accretion torque}\label{SubSec:TROm}

The time-resolved $Q(t)$ analysis in Section \ref{SubSec:QVsFL} can be generalized to other state variables to gain additional insight into the relative performance of the models ${\cal M}_{\rm PP}$ and ${\cal M}_{\rm RP}$. In this section, we examine the magnetocentrifugal accretion torque $I\dot{\Omega}(t)$ --- or, equivalently, the angular acceleration $\dot{\Omega}(t)$ --- inferred by the Kalman filter in Appendix \ref{App:KF_PE}. The goal is to compare the fluctuations in $Q(t)$ and $\dot{\Omega}(t)$, which are related by $\dot{\Omega}(t) \propto Q(t)$ through Equation (\ref{eq:torque}). Spectral and X-ray timing analyses of BATSE, Swift/BAT, and Fermi $\nu(t)$ observations of 4U 1626$-$67 near MJD 48000 (1990 torque
reversal from acceleration to deceleration) and MJD 54500 (2008 torque reversal from deceleration to acceleration) reveal that the absolute value of $\dot{\Omega}(t) \propto Q(t) R_{\rm m}(t)^{1/2} n[\omega(t)]$ shifts by $\lesssim 20\%$ on either side of the reversals \citep{Wilson_1993,Bildsten_1994,Bildsten_1997,Chakrabarty_19974U,Camero_2009}.  For example, \cite{Bildsten_1994} inferred $\dot{\Omega} \approx 5.3 \times 10^{-12} \, \rm{rad \, s^{-2}}$ during acceleration and $\dot{\Omega} \approx -4.6 \times 10^{-12} \, \rm{rad \, s^{-2}}$ during deceleration for the 1990 torque reversal near MJD 48000, and \cite{Camero_2009} inferred $\dot{\Omega} \approx -3.0 \times 10^{-12} \, \rm{rad \, s^{-2}}$ during deceleration and $\dot{\Omega} \approx 2.5 \times 10^{-12} \, \rm{rad \, s^{-2}}$ during acceleration for the 2008 torque reversal near MJD 54500.

In Figure \ref{fig4} we plot the time-resolved angular acceleration $\dot{\Omega}(t_{n'})$ (units: $\rm{rad \, s^{-2}}$; magenta points, left column) versus $t_{n'}$ (units: MJD) and $\dot{\Omega}(t_{n''})$ (units: $\rm{rad \, s^{-2}}$; cyan points, right column) versus $t_{n''}$ (units: MJD) assuming $\mathcal{M}_{\rm PP}$ [$\epsilon(t_{n'}) =  \epsilon(t_{n''}) = 1$; top row] and $\mathcal{M}_{\rm RP}$ [$\epsilon(t_{n'}) = -1$ during deceleration and $\epsilon(t_{n''}) = 1$ during acceleration; bottom row], with $1 \leq n' \leq N_{\rm B}$ and $N_{\rm B} + 1 \leq n'' \leq N_{\rm B} + N_{\rm F}$. Each plotted point corresponds to one $\dot{\Omega}(t_n)$ sample. That is, we evaluate $\dot{\Omega}(t_n) = Q(t_n) [GM \, R_{\rm m}(t_n)]^{1/2} n[\omega(t_n)]/I$ for every state estimate $\Omega(t_n)$, $Q(t_n)$, and $S(t_n)$ returned by the Kalman filter in Appendix \ref{App:KF_PE}, with $1 \leq n \leq N = N_{\rm B} + N_{\rm F}$. The top right and bottom right panels are identical, because
$\mathcal{M}_{\rm PP}$ and $\mathcal{M}_{\rm RP}$ both assume, that the disk is prograde during the acceleration episode
(right column).

The results in Figure \ref{fig4} display two key features. First, $\dot{\Omega}(t_{n'})$ satisfies $-6 \lesssim \dot{\Omega}(t_{n'})/(10^{-11} \, \rm{rad \, s^{-2}}) \lesssim -4$ and $2 \lesssim \dot{\Omega}(t_{n''})/(10^{-12} \, \rm{rad \, s^{-2}}) \lesssim 9$ assuming $\mathcal{M}_{\rm PP}$ and $-9 \lesssim \dot{\Omega}(t_{n'})/(10^{-12} \, \rm{rad \, s^{-2}}) \lesssim -5$ and $2 \lesssim \dot{\Omega}(t_{n''})/(10^{-12} \, \rm{rad \, s^{-2}}) \lesssim 9$ assuming $\mathcal{M}_{\rm RP}$. Second, looking at the two panels in the top row, which display $\dot{\Omega}(t_n)$ associated with $\mathcal{M}_{\rm PP}$, we observe an abrupt decrease in $\dot{\Omega}(t_{n})$ from $|\dot{\Omega}(t_{n'})| \approx 4.0 \times 10^{-11} \, \rm{rad \, s^{-2}}$ near MJD 51500 to $|\dot{\Omega}(t_{n''})| \approx 3.5 \times 10^{-12} \, \rm{rad \, s^{-2}}$ near MJD 55000. In contrast, the two panels in the bottom row reveal, that $\dot{\Omega}(t_n)$ associated with $\mathcal{M}_{\rm RP}$ transitions relatively smoothly on either side of the torque reversal from $|\dot{\Omega}(t_{n'})| \approx 6.0 \times 10^{-12} \, \rm{rad \, s^{-2}}$ near MJD 51500 to $|\dot{\Omega}(t_{n''})| \approx 3.5 \times 10^{-12} \, \rm{rad \, s^{-2}}$ near MJD 55000. The foregoing results for $\mathcal{M}_{\rm RP}$ are broadly consistent with previous results from X-ray timing and spectral analyses  \citep{Wilson_1993,Bildsten_1994,Chakrabarty_19974U, Camero_2009}, adding to the indirect, circumstantial evidence favoring $\mathcal{M}_{\rm RP}$ over $\mathcal{M}_{\rm PP}$ (see Section \ref{SubSec:QVsFL}).
\begin{figure}
\centering{
    \includegraphics[width=\textwidth, keepaspectratio]{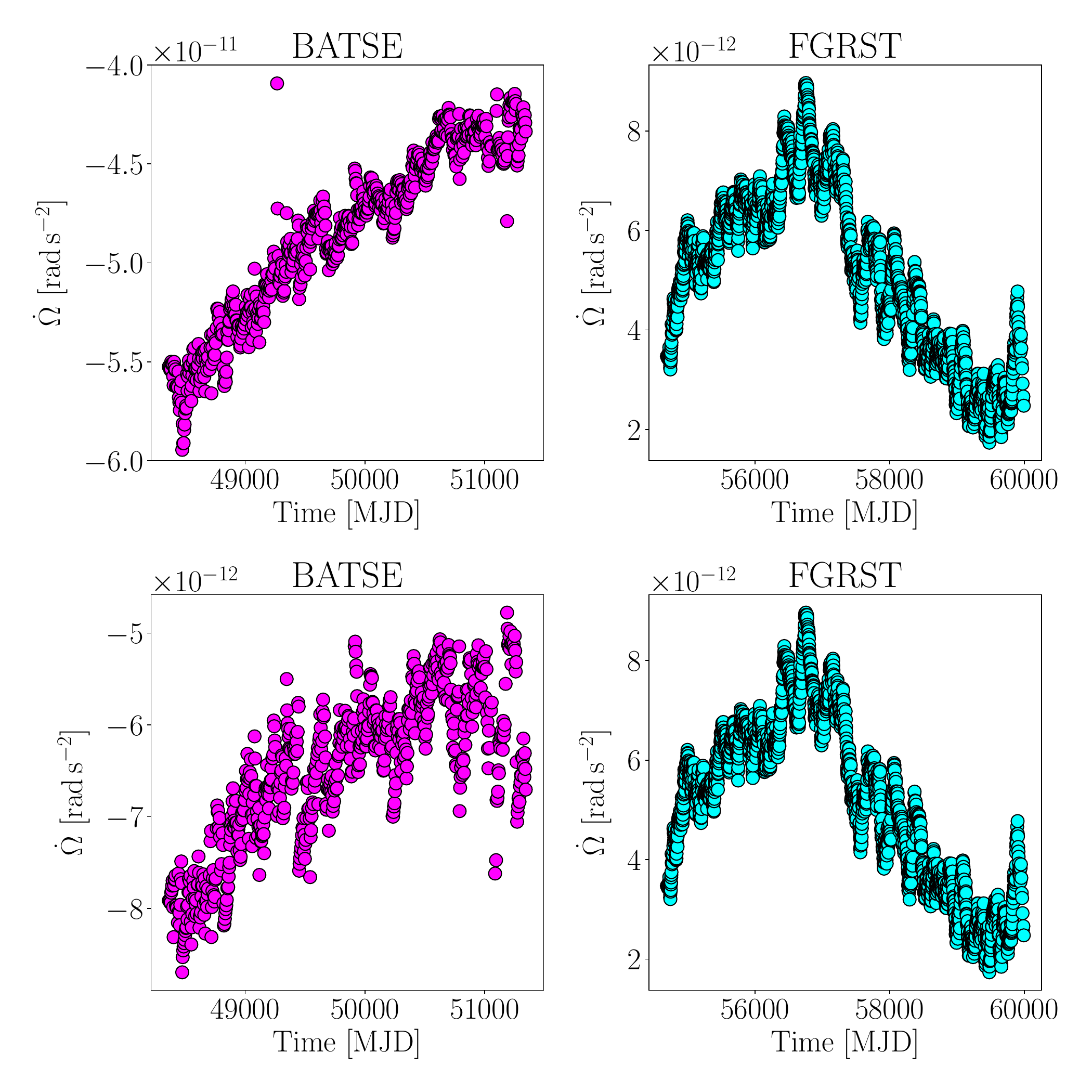}}
    \caption{Time-resolved angular acceleration histories $\dot{\Omega}(t_{n'})$ (units: ${\rm rad \, s^{-2}}$; BATSE observations; magenta points, left column) versus $t_{n'}$ (units: MJD)  and $\dot{\Omega}(t_{n''})$ (units: ${\rm rad \, s^{-2}}$; FGRST observations; cyan points, right column) versus $t_{n''}$ (units: MJD) assuming $\mathcal{M}_{\rm PP}$ [top row; $\epsilon(t_{n'}) = \epsilon(t_{n''}) = 1$ during deceleration (left column) and acceleration (right column)] and $\mathcal{M}_{\rm RP}$ [bottom row; $\epsilon(t_{n'}) = -1$ during deceleration (left column) and $\epsilon(t_{n''}) = 1$ during acceleration (right column)] respectively, with $1 \leq n' \leq N_{\rm B}$ and $N_{\rm B} + 1 \leq n'' \leq N_{\rm B} + N_{\rm F}$. The top right and bottom right panels are identical, because ${\cal M}_{\rm PP}$ and ${\cal M}_{\rm RP}$ both assume, that the disk is prograde during the acceleration episode (right column). The horizontal and vertical axes are plotted on linear scales.}
    \label{fig4}
\end{figure}

\subsection{Time-resolved fastness $\omega(t)$}\label{SubSec:HiddenState}

The time-dependent fastness $\omega(t)$ controls the sign of the magnetocentrifugal torque through Equation (\ref{eq:torque}). In the traditional picture of magnetocentrifugal accretion, i.e.\ an always-prograde disk with $\epsilon(t) = 1$ for all $t$, one expects $\omega(t) \gtrsim 1$ during deceleration, i.e.\ in the propeller phase (BATSE observations; magenta points, Figure \ref{fig1}), and $0 \lesssim \omega(t) \lesssim 1$ during acceleration, i.e.\ in the accretion phase (FGRST observations; cyan points, Figure \ref{fig1}). In the generalized picture of magnetocentrifugal accretion in Section \ref{Sec:MagDyn}, i.e.\ a retrograde-prograde transition with $\epsilon(t) = -1$ during deceleration (BATSE observations; magenta points, Figure \ref{fig1}) and $\epsilon(t) =1$ during acceleration (FGRST observations; cyan points, Figure \ref{fig1}), one expects $0 \lesssim \omega(t) \lesssim 1$ for all $t$. 

In this section, we seek to distinguish between the two pictures above by tracking $\omega(t)$ as a function of $t$ across the 2008 torque reversal. We apply the Kalman filter and nested sampler in Appendix \ref{App:KF_PE} to the $N = N_{\rm B} + N_{\rm F} = 3340$  BATSE and FGRST samples in Figure \ref{fig1}. We employ the MAP estimate of $\bm{\Theta}$ returned by the nested sampler as input to the Kalman filter to generate the time series $\omega(t_1), \hdots \omega(t_{N_{\rm B}})$ and $\omega(t_{N_{\rm B}+1}), \hdots \omega(t_{N_{\rm B} + N_{\rm F}})$; see Sections 3 and 4 in \cite{OLeary_2024b} and \cite{OLeary_2024a} for details of similar analyses. We evaluate $\omega(t_n) = [R_{\rm m}(t_n)/R_{\rm c}(t_n)]^{3/2}$ for every $\Omega(t_n)$, $Q(t_n)$, and $S(t_n)$ returned by the Kalman filter, where $R_{\rm m}(t_n)$ and $R_{\rm c}(t_n)$ are defined in terms of the magnetospheric variables in Section \ref{Sec:MagDyn}, for $1\leq n \leq N$. 

In Figure \ref{fig5} we plot $\omega(t_{n'})$ versus $t_{n'}$ (magenta points, left column) and $\omega(t_{n''})$ versus $t_{n''}$ (cyan points, right column) assuming $\mathcal{M}_{\rm PP}$ (top row) and $\mathcal{M}_{\rm RP}$ (bottom row), with $1 \leq n' \leq N_{\rm B}$ and $N_{\rm B} + 1 \leq n'' \leq N_{\rm B} + N_{\rm F}$. To guide the physical interpretation, we draw three black, dashed, horizontal lines to divide each panel into the ordered unstable [$\omega(t) \lesssim 0.45$], chaotic unstable [$0.45 \lesssim \omega(t) \lesssim 0.6$], and stable [$0.6 \lesssim \omega(t) \lesssim 1.0$] accretion regimes identified by numerical simulations \citep{Blinova_2016}. The weak propeller regime occurs for $1 \lesssim \omega(t) \lesssim 1.25$ \citep{Spruit_1993,Dangelo_2010,Papitto_2015}, so we extend the vertical axes of all panels to span $0 \leq \omega(t) \leq 1.25$, except for the top left panel, where we find $\omega(t_{n'}) > 1.25$ for all $t_{n'}$. In the top panel, we extend the vertical axis  to $0 \leq \omega(t_{n'}) \leq 4.5$ ($1 \leq n' \leq N_{\rm B}$). 

The results in Figure \ref{fig5} have three key features. First, during deceleration (magenta points, left column), we observe that $\omega(t_{n'})$ remains in the strong propeller regime, if one assumes a prograde disk [$\epsilon(t)=1$]. In contrast, it remains in the ordered unstable accretion regime, if one assumes a retrograde disk [$\epsilon(t)=-1$]. Second, upon visual inspection of the two panels in the top row, which display the fastness $\omega(t_n)$ associated with $\mathcal{M}_{\rm PP}$, we observe an abrupt shift in $\omega(t_n)$ from $\omega(t) \approx 3.0$ near MJD 51500 to $\omega(t_n) \approx 0.30$ near MJD 55000. In contrast, visual inspection of the two panels in the bottom row reveals that the fastness $\omega(t_n)$ associated with $\mathcal{M}_{\rm RP}$ transitions relatively smoothly on either side of the torque reversal from $\omega(t) \approx 0.25$ near MJD 51500 to $\omega(t) \approx 0.30$ near MJD 55000. Third, it is interesting physically that the fastness $\omega(t_{n''})$ ($N_{\rm B} + 1 \leq n'' \leq N_{\rm B} + N_{\rm F}$) associated with the FGRST acceleration data (cyan points, right column) spends $\approx 25\%$ of the total observation time visiting the chaotic unstable accretion regime. 

It is unclear astrophysically, whether the weak and strong propeller regimes are separated by a sharp boundary at $\omega(t) = 1.25$ \citep{Spruit_1993,Dangelo_2010,Papitto_2015}. Intuitively, however, one expects some accretion to occur for $1 \lesssim \omega(t) \lesssim 1.25$. Specifically, in the weak propeller regime, enough angular momentum is deposited in the inner disk (due to the complex disk-magnetosphere interaction) to inhibit accretion, but not enough to centrifugally expel material from the disk and produce a vertical outflow \citep{Matt_2005,Melatos_2022}. The accumulated gas at the edge of the inner disk eventually breaks through the magnetocentrifugal barrier, e.g.\ due to hydromagnetic instabilities at the disk-magnetosphere boundary, so that accretion onto the star occurs in episodic bursts \citep{Spruit_1993,Dangelo_2010,Melatos_2022}. In the strong propeller regime $\omega(t) > 1.25$, in contrast, persistent large-scale outflows are expected to occur, with most of the gas at the inner disk being expelled centrifugally, as revealed by magnetohydrodynamic simulations \citep{Romanova_2018}. Hence it is challenging to reconcile the BATSE $\nu(t)$ observations, visible in Figure \ref{fig1} as magenta points, with the results in the top left panel of Figure \ref{fig5}, if one postulates a prograde disk during deceleration. That is, the BATSE observations in Figure \ref{fig1} suggest that 4U 1626--67 accretes persistently, while the fastness results in Figure \ref{fig5}, i.e.\ $\omega(t_{n'})>1.25$ $(1 \leq n'\leq N_{\rm B})$, suggest that most of the material deposited at the inner edge of the disk should be expelled. In contrast, if one assumes a retrograde disk during deceleration, the results in Figures \ref{fig2} and \ref{fig5} are compatible.
 
 \begin{figure}
\centering{
    \includegraphics[width=\textwidth, keepaspectratio]{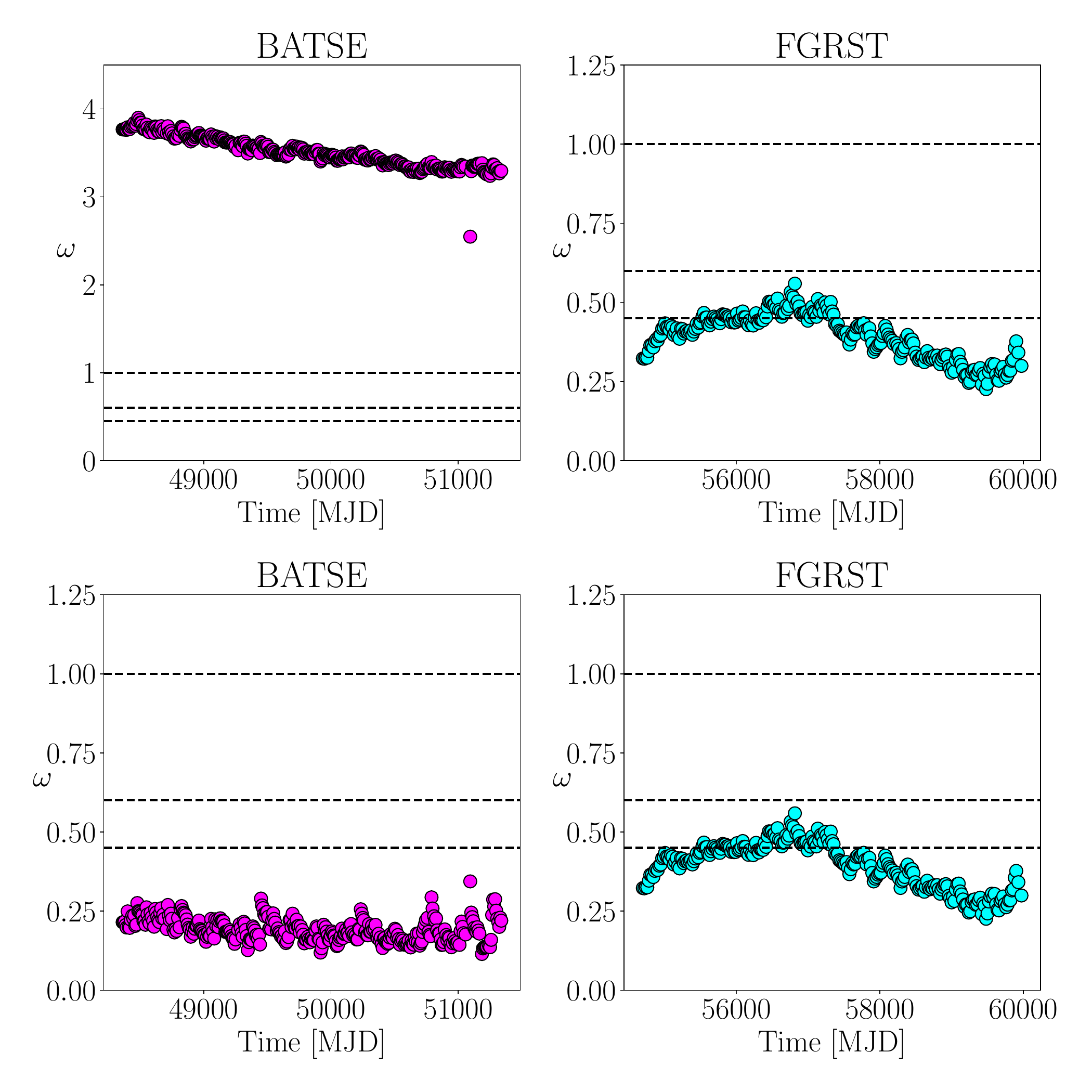}}
    \caption{Time-resolved fastness histories $\omega(t_{n'})$ (BATSE observations; magenta points, left column) versus $t_{n'}$ (units: MJD)  and $\omega(t_{n''})$ (FGRST observations; cyan points, right column) versus $t_{n''}$ (units: MJD) assuming $\mathcal{M}_{\rm PP}$ (top row) and $\mathcal{M}_{\rm RP}$ (bottom row) respectively, with $1 \leq n' \leq N_{\rm B}$ and $N_{\rm B} + 1 \leq n'' \leq N_{\rm B} + N_{\rm F}$. We divide each panel into the ordered unstable [$\omega(t) \lesssim 0.45$], chaotic unstable [$0.45 \lesssim \omega(t) \lesssim 0.6$], and stable [$0.6 \lesssim \omega(t) \lesssim 1.0$] accretion regimes identified by numerical simulations \citep{Blinova_2016} using three, black, dashed horizontal lines.  We extend the vertical axes in all panels to accommodate departures into the weak propeller regime $0 \leq \omega(t) \leq 1.25$ (all panels except the top left) and the strong propeller regime $1.25 \leq \omega(t) \leq 4.5$ (top left panel). The top right and bottom right panels are identical, because ${\cal M}_{\rm PP}$ and ${\cal M}_{\rm RP}$ both assume, that the disk is prograde during the acceleration episode (right column).}
    \label{fig5}
\end{figure}

\section{Conclusion}\label{Sec:Conclusion}
In this paper, we apply the Kalman filter framework developed by \cite{Melatos_2022} to $N = N_{\rm B} + N_{\rm F} = 3340$ BATSE and FGRST observations of the persistent, Galactic, accretion-powered pulsar 4U 1626--67 bracketing the 2008 torque reversal \citep{Camero_2009,Jain_2010,Camero_2012,Beri_2014,turkouglu_2017,Benli_2020,Genccali_2022,Serim_2023}. We modify Equation (\ref{eq:torque}) with a prefactor $\epsilon(t)$, to allow for either a prograde [$\epsilon(t) = 1$] or retrograde [$\epsilon(t)=-1$] disk and test the standard picture of magnetocentrifugal accretion for consistency using two models, namely $\mathcal{M}_{\rm PP}$ and $\mathcal{M}_{\rm RP}$. The analysis tracks $Q(t)$, $\dot{\Omega}(t)$, and $\omega(t)$ independently and sheds new light onto the complex accretion physics near the disk-magnetosphere boundary during the 2008 torque reversal in 4U 1626$-$67. 

The key findings of the analysis are numbered and summarized below.
\begin{enumerate}
    \item[i.] We infer a significant (albeit moderate) preference for the retrograde-prograde model $\mathcal{M}_{\rm RP}$, with log Bayes factor $\log_{10} \mathcal{B} = 0.44$ and MAP Kalman log likelihood ratio $\log_{10} \Lambda = 2.5$. A Bayes factor $\log_{10} {\cal B} \gtrsim 0.5$ is arbitrarily yet widely regarded as ``substantial evidence'' for ${\cal M}_{\rm RP}$ over ${\cal M}_{\rm PP}$ \citep{Kass_1995,Jeffreys_1998}.
    \item[ii.] We infer $\bar{Q} = 3.0^{+0.13}_{-0.11} \times 10^{16} \, \rm{g \, s^{-1}}$ during deceleration and $\bar{Q} = 1.5^{+0.28}_{-0.21} \times 10^{16} \, \rm{g \, s^{-1}}$ during acceleration for the retrograde-prograde model $\mathcal{M}_{\rm RP}$, in accord with $\bar{Q}$ values deduced observationally \citep{Chakrabarty_19974U,Chakrabarty_1998,Sharma_2023,Tobrej_2024} or otherwise \citep{daiAnt_2017,Schulz_2019,Benli_2020}. The peaks of the one-dimensional $\bar{Q}$ posteriors shift by $\approx 0.34$ dex on either side of the transition, indicating that $\bar{Q}$ transitions relatively smoothly through the 2008 torque reversal. 
    
    \item[iii.] We find that the time-resolved mass accretion rate $Q(t)$ associated with $\mathcal{M}_{\rm RP}$ shares the same qualitative features as the aperiodic X-ray flux $F_{X}(t)$ during deceleration. The fluctuations revert to the mean on a timescale of $\sim \, \rm{years}$,  with $\gamma_{Q,{\rm RP}} = 1.1^{+0.23}_{-0.10} \times 10^{-8} \, \rm{s^{-1}}$.
    
    \item[iv.] The angular acceleration $\dot{\Omega}(t)$ output by the Kalman filter satisfies $-9 \lesssim \dot{\Omega}(t_{n'})/(10^{-12} \, \rm{rad \, s^{-2}}) \lesssim -5$ during deceleration and $2 \lesssim \dot{\Omega}(t_{n''})/(10^{-12} \, \rm{rad \, s^{-2}}) \lesssim 9$ during acceleration assuming $\mathcal{M}_{\rm RP}$, broadly consistent with measurements by \cite{Camero_2009}, viz.\ $\dot{\Omega} \approx -3.0 \times 10^{-12} \, \rm{rad \, s^{-2}}$ and $\dot{\Omega} \approx 2.5 \times 10^{-12} \, \rm{rad \, s^{-2}}$.
    \item[v.] In the model $\mathcal{M}_{\rm RP}$, the time-resolved fastness transitions smoothly across the 2008 torque reversal from $\omega(t) \approx 0.25$ near MJD 51500 to $\omega(t) \approx 0.30$ near MJD 55000. In the model ${\cal M}_{\rm PP}$, the fastness switches abruptly from $\omega(t) \approx 3.0$ near MJD 51500 to $\omega(t) \approx 0.30$ near MJD 55000. At the time of writing there is no compelling astrophysical mechanism to support such an abrupt shift in $\omega(t)$ \citep{Nelson_1997,Yi_1997,Yi_1999,Dai_2006}, lending further indirect support to $\mathcal{M}_{\rm RP}$ over $\mathcal{M}_{\rm PP}$. We emphasize, however, that an abrupt shift is not ruled out.
 
\end{enumerate}

The results in points (i)--(v) above are encouraging and provide the first indirect time-resolved evidence supporting a retrograde accretion disk around 4U 1626$-$67 during deceleration. However, we emphasize that they do not prove by themselves that there is a retrograde disk around 4U 1626$-$67. Independent corroborating evidence is required to make sure. Points (i)--(v) are examples of the new and astrophysically important information that stems from a time-resolved Kalman filter analysis of X-ray timing observations of accretion-powered pulsars. There are opportunities to leverage the Kalman filter framework in this paper to deepen the contact between theory and observations, as more data become available. 

This research was supported by the Australian Research Council Centre of Excellence for Gravitational Wave Discovery, grant number CE170100004. NJO’N is the recipient of a Melbourne Research Scholarship. DMC acknowledges funding through the National Science Foundation Astronomy and astrophysics research grant 2109004.  DMC, SB, and STGL acknowledge funding through the National Aeronautics and Space Administration Astrophysics Data Analysis Program grant NNX14-AF77G.

\appendix 
\section{Kalman filter parameter estimation}\label{App:KF_PE}
The goal of recursive Bayesian estimation is to sequentially compute the posterior density of the hidden (``latent'') state variables $\bm{X}(t_n)$, given a sequence of observations $\bm{Y}(t_n)$ and a set of static model parameters $\bm{\Theta}$, for $1\leq n \leq N$. Here we employ a nested sampler \citep{Skilling_2004,Skilling_2006} to find the value of $\bm{\Theta}$ that maximizes the Kalman filter likelihood, namely the MAP estimate, subject to suitable astrophysical priors $p(\bm{\Theta})$ and conditional on $\bm{Y}(t_1), \hdots, \bm{Y}(t_N)$. The reader is referred to: (i) \cite{OLeary_2024a} for a detailed practical guide on applying the framework of \cite{Melatos_2022} to real astronomical data; (ii) Section 3 and Appendix A of \cite{OLeary_2024b} for summaries of the nested sampling and Kalman filter algorithms employed in this paper; and (iii) \cite{Meyers_2021}, \cite{ONeill_2024}, and  \cite{Kimpson_2024a,Kimpson_2024b, kimpson_2025} for overviews of related but different parameter estimation problems in pulsar astrophysics, namely tracking radio pulsar timing noise and detecting continuous gravitational waves with a pulsar timing array, respectively.

\subsection{Recursion relations and likelihood}\label{SubSec:Recursion}

Given a pulse frequency time series $\bm{Y} = [\nu(t_n)]$ with $1 \leq n \leq N$, the unscented Kalman filter employed in this paper sequentially estimates three quantities: the hidden state $\bm{X}_{n} = [\Omega(t_n), Q(t_n), S(t_n)]$, the hidden state error covariance $\bm{P}_{n}$, and the likelihood $\mathcal{L} = p(\{\bm{Y}\}_{n=1}^{N}| \bm{\Theta})$, assumed here to be Gaussian. The recursive updates of $\bm{X}_n$ and $\bm{P}_{n}$ depend on the respective dynamical and measurement uncertainties, quantified by the process and measurement noise covariances, denoted by $\bm{Q}_n$ and $\bm{\Sigma}_n$ at time $t_n$, respectively. With respect to $\mathcal{L}$, the Kalman filter computes the expected value of the observables $\bm{Y}^{-}_{n}$, defined implicitly by the noiseless term on the right-hand side of Equation (\ref{eq:nu}). The measurement residual $\bm{e}_n = \bm{Y}_n - \bm{Y}^{-}_{n}$ and associated covariance $\langle \bm{e}_n \, \bm{e}_n^{\rm{T}} \rangle$ are then used to calculate $\mathcal{L}$, where the superscript `T' denotes the matrix transpose. For brevity, we do not write out the Kalman recursion equations explicitly, nor their numerous auxiliary quantities, as this is done elsewhere. Specifically, we refer the reader to Appendices A and B of \cite{OLeary_2024b} for more details about the unscented Kalman filter algorithm \citep{Julier_1997,Wan_2000,Wan_2001} and a summary of its output, respectively. 

\subsection{Inference outputs} \label{SubSec:InferenceOutputs}
We employ the \texttt{dynesty} nested sampler \citep{Speagle_2020} through the \texttt{bilby} front end \citep{Ashton_2019} using the \texttt{JointLikelihood} \texttt{bilby} class\footnote{\href{https://lscsoft.docs.ligo.org/bilby/api/bilby.core.likelihood.JointLikelihood.html}{https://lscsoft.docs.ligo.org/bilby/api/bilby.core.likelihood.JointLikelihood.html}}  to infer the posterior distributions $p(\bm{\Theta}| \bm{Y}_{\rm{B}},\bm{Y}_{\rm{F}}, \mathcal{M}_{\rm{RP}})$ and $p(\bm{\Theta}| \bm{Y}_{\rm{B}},\bm{Y}_{\rm{F}}, \mathcal{M}_{\rm{PP}})$, the output of which is discussed in Sections \ref{SubSec:MagParams}. The subscripts `B' and `F' label the static model parameters $\bm{\Theta} = \bm{\Theta}_{\rm{B}} \cup \bm{\Theta}_{\rm{F}}$ and measured time series $\bm{Y}_{\rm{B}}$ and $\bm{Y}_{\rm{F}}$ associated with the BATSE (`B') deceleration (magenta points, Figure \ref{fig1}) and FGRST (`F') acceleration (cyan points, Figure \ref{fig1}) episodes, respectively. As well as the foregoing posterior distributions, the analysis also yields the MAP point estimates $\bm{\Theta}_{\rm{RP,MAP}} = \argmax_{\bm{\Theta}} p(\bm{\Theta}| \bm{Y}_{\rm{B}},\bm{Y}_{\rm{F}}, \mathcal{M}_{\rm{RP}}) $ and $\bm{\Theta}_{\rm{PP, MAP}} = \argmax_{\bm{\Theta}} p(\bm{\Theta}| \bm{Y}_{\rm{B}},\bm{Y}_{\rm{F}}, \mathcal{M}_{\rm{PP}})$, as well as the Kalman filter joint likelihood and Bayesian evidence $\mathcal{Z}$, denoted by $p(\bm{Y}_{\rm{B}},\bm{Y}_{\rm{F}}|\bm{\Theta}_{\rm{RP, MAP}}, \mathcal{M}_{\rm{RP}})$ and $\mathcal{Z}(\bm{Y}_{\rm{B}}, \bm{Y}_{\rm{F}}| \mathcal{M}_{\rm{RP}})$ and $p(\bm{Y}_{\rm{B}},\bm{Y}_{\rm{F}}|\bm{\Theta}_{\rm{PP, MAP}}, \mathcal{M}_{\rm{PP}})$ and  $\mathcal{Z}(\bm{Y}_{\rm{B}}, \bm{Y}_{\rm{F}}| \mathcal{M}_{\rm{PP}})$, respectively. 

The evidence returned by the nested sampler is the probability of $\bm{Y}$ given $\mathcal{M}$. It offers a way to compare $\mathcal{M}_{\rm{RP}}$ and $\mathcal{M}_{\rm{PP}}$, viz. $\log_{10} \mathcal{B} = \log_{10} \mathcal{Z}(\bm{Y}_{\rm{B}}, \bm{Y}_{\rm{F}}| \mathcal{M}_{\rm{RP}})  - \log_{10} \mathcal{Z}(\bm{Y}_{\rm{B}}, \bm{Y}_{\rm{F}}| \mathcal{M}_{\rm{PP}})$. Similarly, the Kalman filter ingests $\bm{\Theta}_{\rm{RP,MAP}}$ and $\bm{\Theta}_{\rm{PP,MAP}}$ as point estimates of the static model parameters $\bm{\Theta}$, so one can also compare the Kalman filter likelihoods, viz. $\log_{10} \Lambda = \log_{10} p(\bm{Y}_{\rm{B}},\bm{Y}_{\rm{F}}|\bm{\Theta}_{\rm{RP, MAP}}, \mathcal{M}_{\rm{RP}}) - \log_{10} p(\bm{Y}_{\rm{B}},\bm{Y}_{\rm{F}}|\bm{\Theta}_{\rm{PP, MAP}}, \mathcal{M}_{\rm{PP}})$. The joint likelihood and Bayesian evidence are estimated via (for example) Equation (15) in \cite{OLeary_2024b} and Equation (16) in \cite{Speagle_2020}, respectively.  

\section{Magnetic moment}\label{App:misalignment}
The magnetic field strength $B(r)$ and hence magnetic dipole moment $\mu$ of accretion-powered pulsars can be measured directly using cyclotron resonant scattering features (CRSFs) \citep{Trumper_1977,Trumper_1978,Makishima_1999,Makishima_2003,Caballero_2012,Revnivtsev_2016,Konar_2017,Maitra_2017,Staubert_2019}. They can be inferred indirectly using (for example) Keplerian or beat frequency models associated with quasiperiodic oscillations (QPO) observed in the light curve power density spectrum  \citep{Alpar_1985,VanDerKlis_1985,Shaham_1987, Sharma_2025, Zhou_2025}. Specifically, one estimates the magnetospheric radius $R_{\rm m}$ assuming (for example) that the QPO frequency corresponds to the Keplerian frequency at $R_{\rm m}$, whereupon $\mu = B(r = R) R^3/2$ can be subsequently inferred at the stellar surface $R = 10 \, \rm{km}$. Examples of such analyses can be found in \cite{Shinoda_1990} and \cite{Kommers_1998} for 4U 1626$-$67,  \cite{Nespoli_2011} for 1A 1118$-$615, \cite{James_2010} for KS 1947$+$300, \cite{James_2011} for 4U 1901$+$ 03, and \cite{Liu_2022} for Cen X$-$3. We refer the reader to Table 1 of \cite{Staubert_2019} for a further 16 examples comparing $\mu$ and $B$ estimates from different accretion torque models with those measured directly using CRSFs.

In addition to the five data products output by the Kalman filter and nested sampler in Sections \ref{SubSec:ModComparison}--\ref{SubSec:HiddenState}, the Kalman filter also yields the one-dimensional, marginalized posterior of $\mu$ \citep{Melatos_2022,OLeary_2024a,OLeary_2024b}. We elect not to present the new results about $\mu$ in Section \ref{Sec:Torque} for the reasons listed below in Appendix \ref{App:SysUncertainties}. Rather, we present them in this appendix to support future (refined) Kalman filter analyses of X-ray timing observations of accretion-powered pulsars. In Appendix \ref{SubSec:MagneticMoment} we present the magnetic dipole moment in terms of the magnetocentrifugal parameters, $\bar{Q}$ and $\bar{S}$, and lay out a recipe to ensure that $\mu$ remains constant (as befits a stellar property) when joining the BATSE and Fermi data streams. In Appendices \ref{SubSec:AlignedRotator} and \ref{App:Misaligned}, we present the one-dimensional, marginalized $\mu$ posteriors associated with $\mathcal{M}_{\rm PP}$ and $\mathcal{M}_{\rm RP}$ assuming an aligned rotator ($\Theta = 0^{\circ}$) and misaligned rotator ($\Theta = 50^{\circ}$). We summarize some of the associated systematic and statistical uncertainties associated with the canonical magnetocentrifugal paradigm in Appendix \ref{App:SysUncertainties}.

\subsection{Magnetic dipole moment from the nested sampler} \label{SubSec:MagneticMoment}

The magnetic dipole moment $\mu$ of an accretion-powered pulsar is not expected to change materially during the 47 yrs of X-ray timing observations described in Section \ref{Sec:Observations} and visualized in Figure \ref{fig1}, although it may evolve on longer timescales $\gtrsim 10^5 \, {\rm yr}$ due to the Ohmic and Hall effects \citep{Hollerbach_2002,Cumming_2004} or magnetic burial \citep{Payne_2004}. We can express $\mu$ in terms of the magnetospheric parameters $\bar{Q}$ and $\bar{S}$ according to \citep{Melatos_2022,OLeary_2024b}
\begin{equation} \label{Eq:CnstMu}
    \mu = 2^{-5/2} \pi^{-7/10} (GM)^{3/5} \bar{Q}^{6/5} \bar{S}^{-7/10}.
\end{equation}
Assuming that $\mu$ is constant during the observations analyzed in Section \ref{Sec:Observations}, we infer from Equation (\ref{Eq:CnstMu}) the relations
\begin{equation}\label{Eq:ScalingQS1}
\bar{S}_{\rm B,PP} = \bar{S}_{\rm F,PP} (\bar{Q}_{\rm B,PP}/\bar{Q}_{\rm F,PP})^{12/7}
\end{equation}
and 
\begin{equation}\label{Eq:ScalingQS2}
\bar{S}_{\rm B,RP} = \bar{S}_{\rm F,RP} (\bar{Q}_{\rm B,RP}/\bar{Q}_{\rm F,RP})^{12/7}
\end{equation}
by matching across both sides of the 2008 torque transition for $\mathcal{M}_{\rm PP}$ and $\mathcal{M}_{\rm RP}$, respectively. The subscripts `B' and `F' in Equations (\ref{Eq:ScalingQS1}) and (\ref{Eq:ScalingQS2}) denote the magnetospheric parameters inferred from the BATSE and FGRST data, respectively. Hence, in the Kalman filter analysis, we replace $\bar{S}$ in Equation (\ref{Eq:CnstMu}) with the right-hand side of Equation (\ref{Eq:ScalingQS1}) for $\mathcal{M}_{\rm PP}$ and the right-hand  side of Equation (\ref{Eq:ScalingQS2}) for $\mathcal{M}_{\rm RP}$.

\subsection{Aligned rotator $\Theta = 0^{\circ}$} \label{SubSec:AlignedRotator}

In Figure \ref{fig6} we plot the inferred $\mu$ posterior associated with $\mathcal{M}_{\rm RP}$ (red histogram) and $\mathcal{M}_{\rm PP}$ (gray histogram). Overplotted as a blue, dashed line is the magnetic dipole moment inferred from the surface magnetic field strength $B \approx 3.0 \times 10^{12}$ G measured via CRSFs \citep{Orlandini_1998}, with $\mu =  B R^3/2 = 1.5 \times 10^{30} \, \rm{G\, cm^3}$. The black, dashed, vertical lines correspond to the approximate lower and upper limits of $\mu$, with $1.0 \lesssim \mu/(10^{30} \, \rm{G \, cm^3})\lesssim 3.0$, estimated by analyzing quasiperiodic oscillations \citep{Orlandini_1998,Camero_2009}. The vertical and horizontal axes are on linear and $\log_{10}$ scales, respectively.

The results in Figure \ref{fig6} have three key features. First, the one-dimensional $\mu$ posteriors are unimodal with clearly defined peaks. The analysis yields $\mu = 1.47^{+0.25}_{-0.10} \times 10^{30} \, \rm{G \, cm^3}$ for $\mathcal{M}_{\rm PP}$ and $\mu = 6.10^{+1.4}_{-1.1} \times 10^{29} \, \rm{G \, cm^3}$ for $\mathcal{M}_{\rm RP}$, where the central values correspond to the posterior median and the uncertainty is quantified using a 68\% credible interval.  Second, only a small segment ($\approx 2\%$ of the total probability) of the $\mu$ posterior of $\mathcal{M}_{\rm RP}$ lies between the black, dashed bounds inferred from QPOs \citep{Orlandini_1998,Camero_2009}. Third, the mode of the one-dimensional $\mu$ posterior of $\mathcal{M}_{\rm PP}$ peaks near $\mu = 1.5 \times 10^{30} \, \rm{G \, cm^3}$ [blue, dashed line; \cite{Orlandini_1998}]. The absolute difference between the mode of the one-dimensional $\mu$ posterior of $\mathcal{M}_{\rm PP}$ and the value of $\mu$ measured via cyclotron resonant scattering features is $\approx 0.0076$ dex.

\begin{figure}
\centering{
    \includegraphics[width=\textwidth, keepaspectratio]{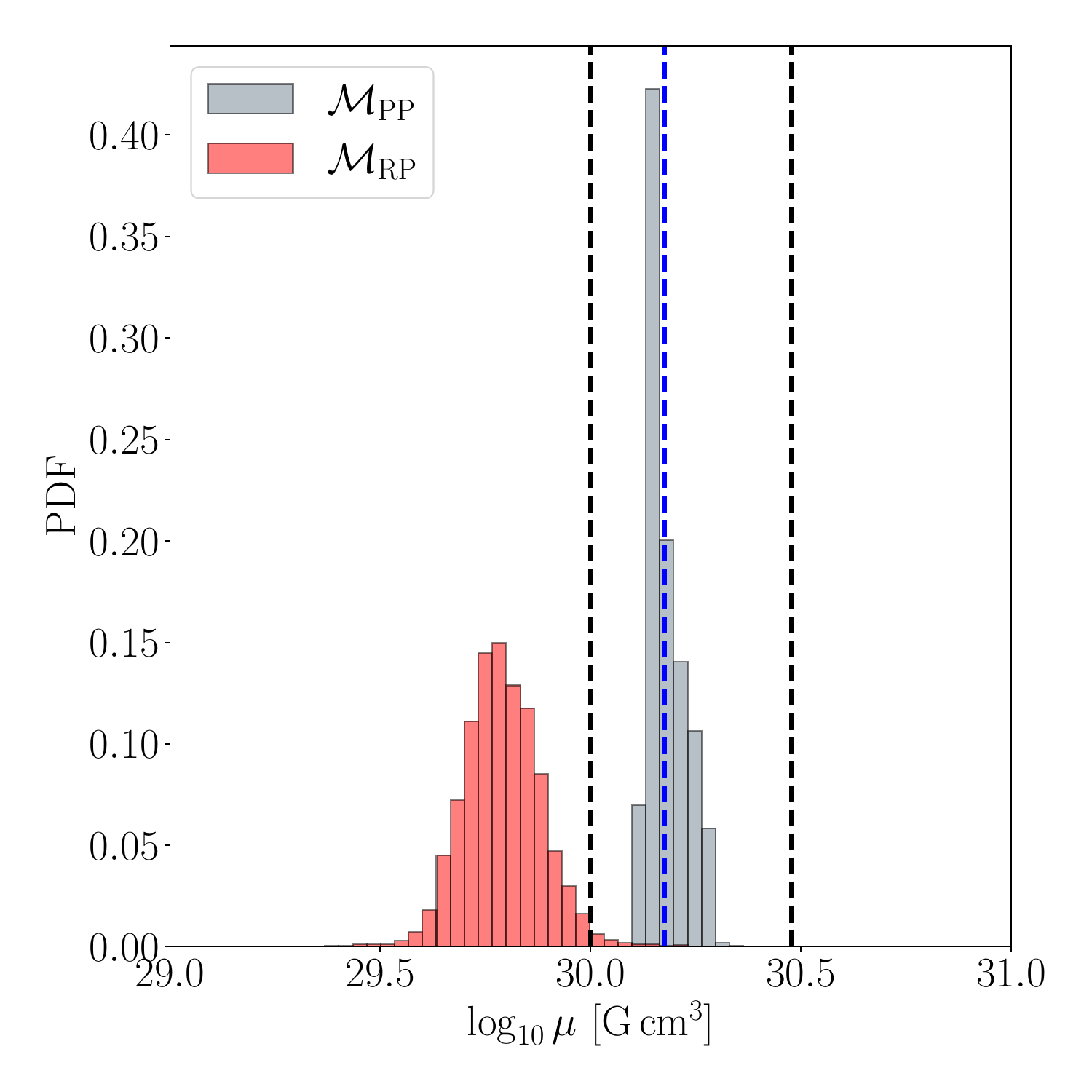}}
    \caption{One-dimensional marginalized posterior distribution of the magnetic dipole moment $\mu$ (units: $\rm{G \, cm^{3}}$) of the persistent X-ray pulsar 4U 1626--67 for $\mathcal{M}_{\rm PP}$ (gray histograms) and $\mathcal{M}_{\rm RP}$ (red histograms) assuming an aligned rotator, i.e.\ $\Theta = 0$. Overplotted as a blue, dashed, vertical line is the $\mu$ estimate of \cite{Orlandini_1998}, measured using cyclotron resonant scattering features. The black, dashed, vertical lines correspond to the approximate lower and upper limits of $\mu$, with $1.0 \lesssim \mu/(10^{30} \, \rm{G \, cm^3})\lesssim 3.0$, inferred from quasiperiodic oscillations \citep{Orlandini_1998,Camero_2009}. The vertical and horizontal axes are on linear and $\log_{10}$ scales respectively. The acronym PDF stands for probability density function.}
    \label{fig6}
\end{figure}

\subsection{Misaligned rotator $\Theta \neq 0^{\circ}$} \label{App:Misaligned}

As a first pass at estimating $\Theta$ for 4U 1626--67, we modify the magnetospheric radius defined in Section \ref{Sec:MagDyn} according to  
\begin{equation}\label{Eq:RmTheta}
         R_{\rm{m}}(t,\Theta) = (2 \pi^{2/5})^{-1} (GM)^{1/5} Q(t)^{2/5} [S(t) f(\Theta)]^{-2/5},
     \end{equation}
 where Equation (\ref{Eq:RmTheta}) coincides respectively with Equations (4) and (6) of \cite{Campana_2001} and \cite{Jetzer_1998} for $f(\Theta) = 1 + 3 \sin^2 \Theta$. At first glance, it may appear that $R_{\rm m}$ compresses, as $\Theta$ increases. However, $S(t) = (2 \pi)^{-1} \mu^2 R_{\rm m}(t,\Theta)^{-6}$ contains $R_{\rm m}(t, \Theta)$ implicitly. Hence, upon substituting the latter expression into Equation (\ref{Eq:RmTheta}), we find that $R_{\rm m}(t,\Theta)$ is smallest for $\Theta = 0$ and greatest for $\Theta = \pi/2$. The magnetospheric radius (for $\Theta \neq 0$) is larger than the minimum magnetospheric radius by a factor of $ (1 + 3 \sin^2 \Theta)^{2/7} \leq 1.49$ \citep{Campana_2001}.

We use Equation (\ref{Eq:RmTheta}) to estimate $\Theta$ by the following rudimentary (and contestable) argument. Let us accept temporarily that ${\cal M}_{\rm RP}$ is preferred over ${\cal M}_{\rm PP}$, as implied by the results in Sections \ref{SubSec:ModComparison} (Bayes factor), \ref{SubSec:MagParams} (smooth transition of $\bar{Q}$), \ref{SubSec:QVsFL} [consistent qualitative features between $Q(t)$ and $F_X(t)$], \ref{SubSec:TROm} [inferred $\dot{\Omega}(t)$ is consistent with previous X-ray timing and spectral analyses \citep{Wilson_1993,Bildsten_1994, Bildsten_1997,Camero_2009}], and \ref{SubSec:HiddenState} [smooth transition of the fastness $\omega(t)$]. Then we can bring the (seemingly contradictory) results concerning $\mu$ in Appendix \ref{SubSec:MagneticMoment} into line with the preference for ${\cal M}_{\rm RP}$ by adjusting $\Theta$, so that the Kalman-filter-inferred $\mu$ value for ${\cal M}_{\rm RP}$ (as opposed to ${\cal M}_{\rm PP}$) matches the independent cyclotron resonant scattering measurement \citep{Orlandini_1998}. In Figure \ref{fig7} we plot the one-dimensional misalignment-corrected magnetic dipole moment $\mu$ posterior distribution associated with $\mathcal{M}_{\rm RP}$ (red histogram) and $\mathcal{M}_{\rm PP}$ (gray histogram). Overplotted as a blue, dashed line is the magnetic dipole moment inferred from the surface magnetic field strength $B \approx 3.0 \times 10^{12}$ G measured via CRSFs \citep{Orlandini_1998}, with $\mu = 1.5 \times 10^{30} \, \rm{G\, cm^3}$. The black, dashed, vertical lines correspond to the approximate lower and upper limits of $\mu$, with $1.0 \lesssim \mu/(10^{30} \, \rm{G \, cm^3})\lesssim 3.0$, estimated by analyzing QPOs \citep{Orlandini_1998,Camero_2009}. The vertical and horizontal axes are on linear and $\log_{10}$ scales, respectively. The analysis reveals that the peak of the red histogram coincides with the blue, dashed line for $\Theta \approx 50^{\circ}$.
 
\begin{figure}
\centering{
    \includegraphics[width=\textwidth, keepaspectratio]{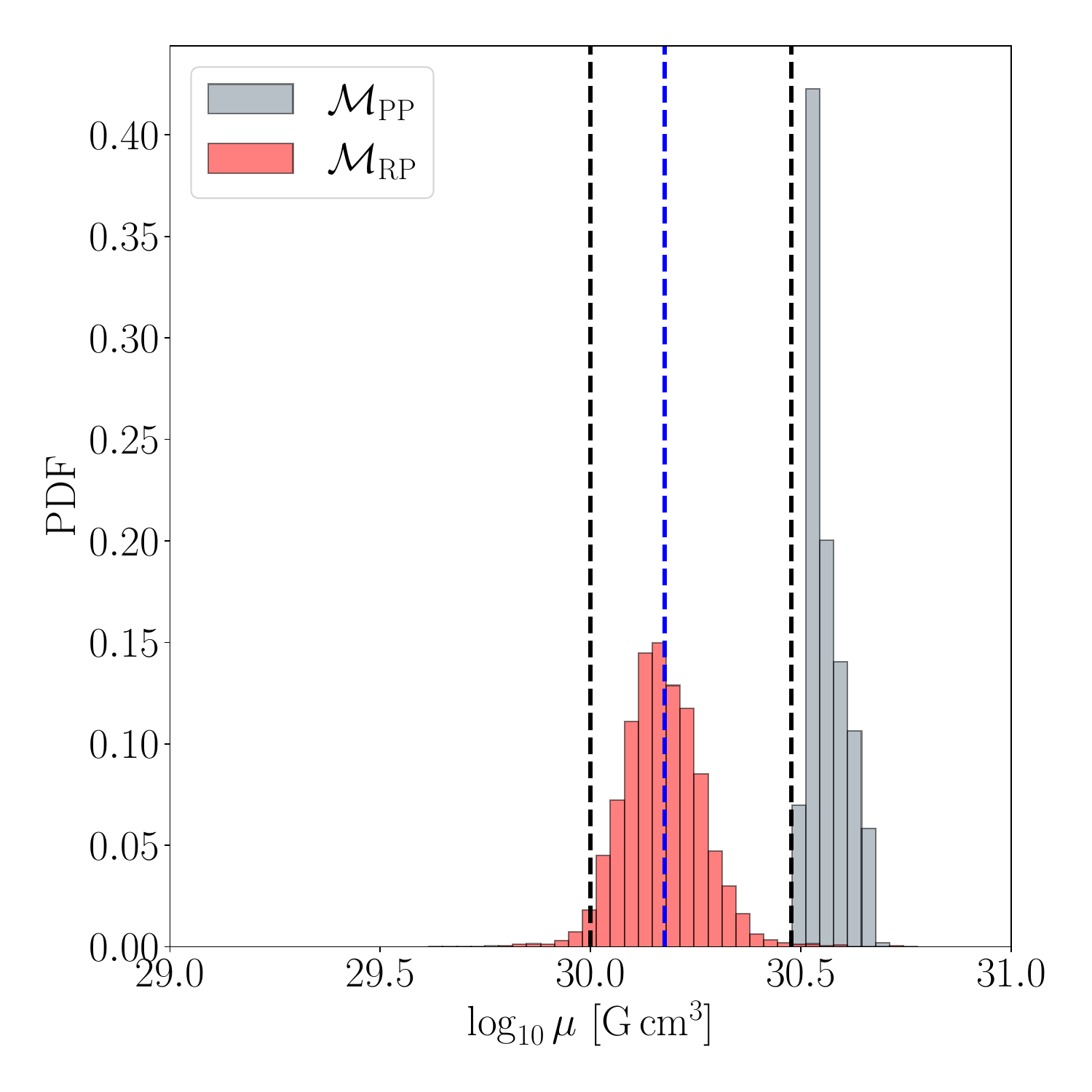}}
    \caption{Same as Figure \ref{fig6} but with $\Theta=50^\circ$ assuming Equation (\ref{Eq:RmTheta}). }
    \label{fig7}
\end{figure}

\subsection{Systematic uncertainties}\label{App:SysUncertainties}

Upon inspecting the results about $\mu$ in Figures \ref{fig6} and \ref{fig7}, it is tempting to conclude that the data favor $\mathcal{M}_{\rm PP}$ over $\mathcal{M}_{\rm RP}$ when $\Theta = 0^{\circ}$ and favor $\mathcal{M}_{\rm RP}$ over $\mathcal{M}_{\rm PP}$ when $\Theta \approx 50^{\circ}$. However, the situation is not clear-cut for the following three reasons.

First, the canonical magnetocentrifugal accretion torque law, Equation (\ref{eq:torque}), supplemented with the mean-reverting dynamics of $Q(t)$ and $S(t)$, Equations (\ref{eq:Lang_Q}) and (\ref{eq:Lang_S}), amount to a highly idealized model in many respects, with systematic uncertainties that should be taken into account when interpreting the $\mu$ estimates. Some of the uncertainties associated with Equations (\ref{eq:torque})--(\ref{eq:Lang_S}) are discussed in Sections 2.2 and 2.3 of \cite{Melatos_2022}, Sections 2.2, 2.3, and 4.4 of \cite{OLeary_2024b}, and Section 2.2 of \cite{OLeary_2024a}. Additional systematic uncertainties arise due to the dimensionless torque prescription, viz.\ $n[\omega(t)]$ \citep{Ghosh_1977,Ghosh_1978,Ghosh_1979,Wang_1995,Rappaport_2004,Matt_2005,Dai_2006,Kluzniak_2007,Gao_2021,Melatos_2022}, the disk geometry, e.g.\ the location of the magnetospheric radius $R_{\rm m} (t)$, the magnetic field geometry, e.g.\ multipolar and misaligned magnetospheres \citep{Makishima_1999,Asseo_2002,Makishima_2003,payne_2006,Suvorov_2020,mushtukov_2023,Tsygankov_2023,suleimanov_2023}, and other important uncertainties related to the aperiodic X-ray flux $F_{X}(t)$ and luminosity $L(t) \propto F_{X}(t)$, e.g.\ outflows \citep{Matt_2005,Matt_2008,Romanova_2015,mushtukov_2023}, X-ray beaming, and bolometric conversion factors. Some of the foregoing systematic uncertainties are discussed in detail in \cite{Takagi_2016} and \cite{Yatabe_2018}.

Second, as well as the above systematic uncertainties, there are additional statistical uncertainties that should be taken into account when interpreting the results about $\mu$. For example, \cite{Christodoulou_2025} demonstrated that deviations of $\mu$, inferred using the Kalman filter in Appendix \ref{App:KF_PE}, are expected by factors of up to four relative to those obtained using CRSFs.

Third, \cite{stierhof_2025} found that  different $n[\omega(t)]$ prescriptions \citep{Ghosh_1977,Ghosh_1978,Ghosh_1979, Wang_1995,Rappaport_2004, Matt_2005, Dai_2006, Kluzniak_2007,Gao_2021} yield estimates of $\mu$ spread over $\sim 1 \, \rm{dex}$; see Figure 8 of \cite{stierhof_2025} for further details. Accordingly, it is challenging to compare fairly the $\mu$ estimates from $\mathcal{M}_{\rm RP}$ and $\mathcal{M}_{\rm PP}$ with those deduced from CRSFs \citep{Orlandini_1998}, without carefully addressing the systematic and statistical uncertainties mentioned in the above two paragraphs, a topic of a forthcoming paper.

\clearpage
\footnotesize
\bibliography{main.bib}{}
\bibliographystyle{aasjournal}

\end{document}